\documentclass[reprint,aps,prb]{revtex4-2}
\usepackage{graphicx}
\usepackage{dcolumn}
\usepackage{bm}
\usepackage{amsmath,amsfonts,amssymb,braket,here}

\begin{document}

\title{Microscopic mechanism for intrinsic nonlinear anomalous Hall conductivity in noncollinear antiferromagnetic metals}

\author{Akimitsu Kirikoshi}
\author{Satoru Hayami}
\affiliation{Department of Physics, Hokkaido University, Sapporo 060-0810, Japan}

\date{\today}

\begin{abstract}
  We theoretically investigate an intrinsic nonlinear anomalous Hall effect (INAHE) in space-time ($\mathcal{PT}$) symmetric antiferromagnetic metals. 
  The INAHE is characterized by an asymmetric and non-dissipative part of the second-order electric conductivity tensor in the clean limit in contrast to the Drude-type symmetric conductivity tensor with dissipation.
  By introducing a multipole description, we show that the emergence of the INAHE is due to active odd-parity magnetic quadrupoles or magnetic toroidal dipoles under magnetic orderings.
  In order to clarify the microscopic origin of the INAHE, we specifically consider a fundamental tight-binding model of a three-dimensional tetragonal system.
  We demonstrate that the INAHE arises from the effective coupling between magnetic ordering and antisymmetric spin--orbit interaction.
  We also discuss essential electron hopping paths driving the INAHE.
\end{abstract}

\maketitle

\section{Introduction}

Intrinsic properties of transport phenomena are important to investigate the functional properties inherent in materials.
One of the prominent examples is the anomalous linear Hall effect \cite{PhysRevLett.83.3737, PhysRevLett.88.207208, JPSJ.71.19, PhysRevLett.93.206602, RevModPhys.82.1539, RevModPhys.82.1959}, whose intrinsic property is accounted for by the Berry curvature in the absence of time-reversal $\mathcal{T}$ symmetry.
The study of transport phenomena has been extended into the second-order nonlinear regime, where the breaking of spatial inversion $\mathcal{P}$ symmetry rather than $\mathcal{T}$ symmetry is necessary~\cite{tokura2018nonreciprocal}.
In the $\mathcal{T}$-symmetric system, the nonlinear conductivity is closely related to the Berry curvature dipole (BCD)~\cite{PhysRevLett.115.216806}, while it is related to an asymmetric band deformation and a quantum metric (QM) tensor in the $\mathcal{T}$-broken system~\cite{PhysRevLett.112.166601, PhysRevLett.124.067203, PhysRevLett.127.277201, PhysRevLett.127.277202}.
Among them, the nonlinear conductivity induced by the QM tensor is intrinsic and independent of the relaxation rate in the clean limit, which might help identify the direction of the N\'{e}el vector in antiferromagnetic (AFM) metals~\cite{PhysRevLett.127.277201, PhysRevLett.127.277202}.

According to the nonlinear response theory based on the Kubo formula, there are mainly three contributions to the second-order electric conductivity: the Drude, BCD, and intrinsic terms~\cite{PhysRevResearch.2.043081, PhysRevX.11.011001, JPSJ.91.014701}.
The Drude term contributes to a symmetric tensor, whereas the BCD and intrinsic terms contribute to an asymmetric tensor and are the origin of the nonlinear Hall effect.
In particular, the latter intrinsic term leads to the intrinsic nonlinear anomalous Hall effect (INAHE) without dissipation.
Through the microscopic model analyses, the key ingredients to cause the Drude-type and BCD-type conductivity have been revealed; the former is induced by an asymmetric band modulation arising from an effective coupling between magnetic ordering and antisymmetric spin--orbit interactions (ASOIs)~\cite{PhysRevResearch.2.043081, PhysRevB.105.155157} or hopping modulations due to an alignment of the local scalar chirality~\cite{PhysRevB.106.014420, JPSJ.91.094704}, whereas the latter is induced by asymmetric hopping paths causing nonzero BCD in momentum space~\cite{PhysRevLett.115.216806, JPSJ.91.014701,kim2019prediction}.
Meanwhile, the microscopic mechanism for intrinsic nonlinear conductivity, including the INAHE, has yet to be fully elucidated.

This paper investigates the INAHE based on symmetry and model analyses.
First, by performing the symmetry analysis, we show that the INAHE is related to the emergence of the magnetic quadrupole and magnetic toroidal dipole, which corresponds to the rank-2 axial and rank-1 polar tensors, respectively.
The former multipole induces the pure INAHE without the Drude-type contribution.
Next, we discuss the microscopic mechanism of the INAHE by analyzing a minimal model in a three-dimensional four-sublattice tetragonal system.
By taking into account electron hoppings and ASOI under two types of noncollinear magnetic orderings accompanying the magnetic quadrupole or magnetic toroidal dipole, we obtain two important ingredients to cause the INAHE:
One is the effective coupling between magnetic order and ASOI, and the other is the closed paths consisting of the nearest-neighbor and further-neighbor hoppings.
We also discuss the difference in the model parameter dependence in the INAHE between the magnetic orderings with magnetic quadrupole and magnetic toroidal dipole.

The organization of this paper is as follows.
Section~\ref{second-order conductivity} introduces the second-order nonlinear conductivity based on the Kubo formula.
To discuss the gauge-invariant quantities, we separate the conductivity tensor into the Ohmic and Hall parts discussed in Ref.~\cite{SciPostPhysCore.5.3.039}.
From the symmetry viewpoint, we also clarify the relationship between the Ohmic/Hall parts and activated multipoles.
Section~\ref{model-calculation} analyzes a minimum model on a layered four-sublattice tetragonal structure under AFM orderings with magnetic quadrupole and magnetic toroidal dipole.
By studying the model parameter dependence and comparing them with the numerical results in each ordered state, we obtain the crucial model parameters and the effective closed paths contributing to the INAHE.
Section~\ref{summary} summarizes this paper and lists the candidate materials to encourage the observations of the pure INAHE based on the magnetic point groups (MPGs).

\section{Second-Order Conductivity}
\label{second-order conductivity}

\subsection{Definition from the second-order Kubo formula}

We first briefly introduce the expression of the nonlinear conductivity to make the present paper self-contained, although its derivation has already been given in the previous literature~\cite{PhysRevResearch.2.043081, JPSJ.91.014701}.
The second-order nonlinear conductivity tensor $\sigma_{\mu;\alpha\beta}$ defined as $j_{\mu}=\sigma_{\mu;\alpha\beta}E_{\alpha}E_{\beta}$ with $\mu,\alpha,\beta=x,y,z$ can be derived from the nonlinear Kubo formula.
In the clean and static limit, the second-order conductivity is classified according to the order of the phenomenological relaxation time $\tau$ as 
\begin{equation}
  \sigma_{\mu;\alpha\beta}=\sigma_{\mu;\alpha\beta}^{\mathrm{D}}+\sigma_{\mu;\alpha\beta}^{\mathrm{BCD}}+\sigma_{\mu;\alpha\beta}^{\mathrm{int}},
  \label{second-order}
\end{equation}
where
\begin{subequations}
  \label{second-order-conductivity}
  \begin{align}
    \sigma^{\mathrm{D}}_{\mu;\alpha\beta}=&\, 
    -\frac{e^{3}\tau^{2}}{2\hbar^{3}V}\sum_{{\bm k}n}f_{n{\bm k}}\partial_{\mu}\partial_{\alpha}\partial_{\beta}\varepsilon_{n{\bm k}},
    \label{Drude}
    \\
    \sigma^{\mathrm{BCD}}_{\mu;\alpha\beta}=&\, 
    \frac{e^{3}\tau}{2\hbar^{2}V}\sum_{{\bm k}n}f_{n{\bm k}}\epsilon_{\mu\alpha\kappa}D_{n}^{\beta\kappa}({\bm k})
    +(\alpha\leftrightarrow\beta),
    \label{BCD}
    \\
    \sigma^{\mathrm{int}}_{\mu;\alpha\beta}=&\, 
    \frac{e^{3}}{\hbar V}\sum_{{\bm k}}\sum_{n,m}^{\varepsilon_{n}\neq\varepsilon_{m}}\left[\frac{1}{2}\frac{f_{n{\bm k}}-f_{m{\bm k}}}{(\varepsilon_{n{\bm k}}-\varepsilon_{m{\bm k}})^{2}}g_{\alpha\beta}^{nm}({\bm k})\partial_{\mu}\varepsilon_{n{\bm k}}\right.
    \notag
    \\
    &\, \left.+2\left(-\frac{\partial f_{n{\bm k}}}{\partial \varepsilon_{n{\bm k}}}\right)\partial_{\alpha}\varepsilon_{n{\bm k}}\frac{g_{\mu\beta}^{nm}({\bm k})}{\varepsilon_{n{\bm k}}-\varepsilon_{m{\bm k}}}\right]
  +(\alpha\leftrightarrow\beta).
  \label{int}
  \end{align}
\end{subequations}
We here adopt a symmetric gauge as $\sigma_{\mu;\alpha\beta}=\sigma_{\mu;\beta\alpha}$.
The first term in Eq.~(\ref{second-order}) represents the Drude term proportional to $\tau^2$, where $\varepsilon_{n{\bm k}}$ and $f_{n{\bm k}}$ are a band energy and a Fermi distribution function with wave vector $\bm{k}$ and band index $n$, respectively.
$e,\hbar$, and $V$ are the elementary charge, the reduced Planck constant, and the system volume, respectively.
The second term in Eq.~(\ref{second-order}) represents the BCD term proportional to $\tau$, where $\epsilon_{\mu\alpha\kappa}$ is the Levi-Civita tensor and $D_{n}^{\beta\kappa}({\bm k})=\partial_{\beta}\Omega_{n}^{\kappa}({\bm k})$ is the BCD with the Berry curvature $\Omega_{n}^{\kappa}({\bm k})$ given by 
\begin{equation}
  \begin{aligned}
    \Omega_{n}^{\kappa}({\bm k})=&\, i\hbar^{2}\sum_{m\neq n}\epsilon_{\kappa\alpha\beta}\frac{v_{\alpha,nm}({\bm k})v_{\beta,mn}({\bm k})}{(\varepsilon_{n{\bm k}}-\varepsilon_{m{\bm k}})^{2}}.
  \end{aligned}
\end{equation}
$v_{\alpha,nm}({\bm k})$ is a Bloch representation for the velocity operator $\hat{v}_{\alpha}({\bm k})=\partial_{\alpha}\hat{h}({\bm k})/\hbar$ [$\hat{h}({\bm k})$ is the ${\bm k}$-resolved Hamiltonian] defined by
\begin{equation}
  v_{\alpha,nm}({\bm k})\equiv\braket{n{\bm k}|\hat{v}_{\alpha}({\bm k})|m{\bm k}}
\end{equation}
with the eigenstates $\ket{n{\bm k}}$ and $\ket{m{\bm k}}$.
The expressions for the Drude and BCD terms coincide with those obtained by the semi-classical Boltzmann formalism~\cite{PhysRevB.104.054429, PhysRevLett.115.216806}.

The third term in Eq.~(\ref{second-order}) that we focus on in the present study represents the intrinsic term independent of  $\tau$, where $g_{\alpha\beta}^{nm}({\bm k})$ is referred to as a QM tensor~\cite{PhysRevLett.112.166601}.
The expression of $g_{\alpha\beta}^{nm}({\bm k})$ is given by 
\begin{equation}
  \begin{aligned}
    g_{\alpha\beta}^{nm}({\bm k})=&\, \frac{\hbar^{2}}{2}\frac{v_{\alpha,nm}({\bm k})v_{\beta,mn}({\bm k})+v_{\beta,nm}({\bm k})v_{\alpha,mn}({\bm k})}{(\varepsilon_{n{\bm k}}-\varepsilon_{m{\bm k}})^{2}}.
  \end{aligned}
\end{equation}
The first term in the square bracket in Eq.~(\ref{int}) indicates the Fermi sea term, while the second term is the Fermi surface term.

The different terms in Eq.~(\ref{second-order}) are distinguished by symmetry. 
In terms of $\mathcal{T}$ symmetry, the Drude and intrinsic terms are $\mathcal{T}$-odd, while the BCD is $\mathcal{T}$-even. 
Thus, $\sigma^{\mathrm{D}}_{\mu;\alpha\beta}$ and $\sigma^{\mathrm{int}}_{\mu;\alpha\beta} $ become nonzero in magnetic ordered states, while $\sigma^{\mathrm{BCD}}_{\mu;\alpha\beta}$ becomes nonzero even in the paramagnetic state.
In addition, as all the terms are $\mathcal{P}$-odd, $\sigma^{\mathrm{BCD}}_{\mu;\alpha\beta}$ vanishes for the $\mathcal{PT}$-symmetric systems as found in the AFM systems.

Their transformation property concerning the point-group symmetry is also different from each other.
To demonstrate that, we decompose $\sigma_{\mu;\alpha\beta}$ into an Ohmic part ($\sigma^{\mathrm{O}}_{\mu;\alpha\beta}$) and a Hall part ($\sigma^{\mathrm{H}}_{\mu;\alpha\beta}$) as follows~\cite{SciPostPhysCore.5.3.039}: 
\begin{equation}
  \sigma_{\mu;\alpha\beta}=\sigma^{\mathrm{O}}_{\mu;\alpha\beta}+\sigma^{\mathrm{H}}_{\mu;\alpha\beta},
\end{equation}
where $\sigma^{\mathrm{O}}_{\mu;\alpha\beta}$ is the symmetric tensor for the interchange of $(\mu \leftrightarrow \alpha,\beta)$ represented by
\begin{align}
  \sigma_{\mu;\alpha\beta}^{\mathrm{O}}=&\, \sigma_{\alpha;\mu\beta}^{\mathrm{O}}=\sigma_{\beta;\alpha\mu}^{\mathrm{O}}.
\end{align}
On the other hand, $\sigma^{\mathrm{H}}_{\mu;\alpha\beta}$ is asymmetric under such an interchange.
Compared to the expressions derived from the Kubo formula in Eqs.~(\ref{second-order-conductivity}), one obtains 
\begin{equation}
  \begin{aligned}
    \sigma^{\mathrm{O}}_{\mu;\alpha\beta}=&\, \sigma_{\mu;\alpha\beta}^{\mathrm{D}}+\sigma_{\mu;\alpha\beta}^{\mathrm{int, O}},
    \\
    \sigma^{\mathrm{H}}_{\mu;\alpha\beta}=&\, \sigma_{\mu;\alpha\beta}^{\mathrm{BCD}}+\sigma_{\mu;\alpha\beta}^{\mathrm{int, H}}. 
  \end{aligned}
\end{equation}
Thus, $\sigma_{\mu;\alpha\beta}^{\mathrm{D}}$ ($\sigma_{\mu;\alpha\beta}^{\mathrm{BCD}}$) contributes to the Ohmic (Hall) part, while $\sigma_{\mu;\alpha\beta}^{\mathrm{int}}$ contributes to both parts, where we denote as $\sigma_{\mu;\alpha\beta}^{\mathrm{int, O}}$ and $\sigma_{\mu;\alpha\beta}^{\mathrm{int, H}}$.
Among them, $\sigma_{\mu;\alpha\beta}^{\mathrm{int, H}}$ corresponds to the INAHE. 
To focus on the behavior of $\sigma_{\mu;\alpha\beta}^{\mathrm{int, H}}$, we suppose the $\mathcal{PT}$-symmetric AFMs in the following discussion, i.e., $\sigma_{\mu;\alpha\beta}^{\mathrm{BCD}}=0$ and $\sigma_{\mu;\alpha\beta}^{\mathrm{int, H}} \neq 0$.

Let us decompose $\sigma_{\mu;\alpha\beta}^{\mathrm{int}}$ in Eq.~(\ref{int}) into $\sigma_{\mu;\alpha\beta}^{\mathrm{int, O}}$ and $\sigma_{\mu;\alpha\beta}^{\mathrm{int, H}}$. 
In the $\mathcal{PT}$-symmetric AFMs, two spin-degenerate bands appear i.e., $\varepsilon_{n{\bm k}}=\varepsilon_{m{\bm k}}$ for a pair of $n\neq m$.
To simplify the expression, we replace the band index $n,m$ as $\nu,\bar{\nu}$ which satisfies $\nu\neq \bar{\nu}$ and $\varepsilon_{\nu\boldsymbol{k}}\neq \varepsilon_{\bar{\nu}\boldsymbol{k}}$.
Then, the summation over $n,m$ is rewritten as 
\begin{equation*}
  \sum_{n,m}^{\varepsilon_{n}\neq \varepsilon_{m}}\to\sum_{\nu,\bar{\nu}}\sum_{n\in \nu}\sum_{m\in\bar{\nu}}.
\end{equation*}
We thereby obtain
\begin{equation}
  \begin{aligned}
    \sigma_{\mu;\alpha\beta}^{\mathrm{int}}
    =&\, \frac{e^{3}}{V}\sum_{{\bm k}}\sum_{\nu,\bar{\nu}}\left[\frac{f_{\nu{\bm k}}-f_{\bar{\nu}{\bm k}}}{(\varepsilon_{\nu{\bm k}}-\varepsilon_{\bar{\nu}{\bm k}})^{2}}v^{\nu}_{\mu}({\bm k})g^{\nu\bar{\nu}}_{\alpha\beta}({\bm k})\right.
    \\
    &\, \left.+2\left(-\frac{\partial f_{\nu{\bm k}}}{\partial \varepsilon_{\nu{\bm k}}}\right)\frac{v^{\nu}_{\beta}({\bm k})g^{\nu\bar{\nu}}_{\mu\alpha}({\bm k})+v^{\nu}_{\alpha}({\bm k})g^{\nu\bar{\nu}}_{\mu\beta}({\bm k})}{\varepsilon_{\nu{\bm k}}-\varepsilon_{\bar{\nu}{\bm k}}}\right],
    \label{int-2}
  \end{aligned}
\end{equation}
where we introduce an interband QM tensor for the bands $\nu,\bar{\nu}$ as 
\begin{equation}
  g^{\nu\bar{\nu}}_{\alpha\beta}({\bm k})\equiv\sum_{n\in\nu}\sum_{m\in\bar{\nu}}g_{\alpha\beta}^{nm}({\bm k})
\end{equation}
and $v_{\mu}^{\nu}({\bm k})=\partial_{\mu}\varepsilon_{\nu{\bm k}}/\hbar$.
Finally, $\sigma_{\mu;\alpha\beta}^{\mathrm{int}}$ in Eq.~(\ref{int-2}) is decomposed into 
\begin{equation}
  \begin{aligned}
    \sigma^{\mathrm{int, O}}_{\mu;\alpha\beta}=&\, \frac{e^{3}}{3V}\sum_{{\bm k}}\sum_{\nu,\bar{\nu}}\left[\frac{f_{\nu{\bm k}}-f_{\bar{\nu}{\bm k}}}{\varepsilon_{\nu{\bm k}}-\varepsilon_{\bar{\nu}{\bm k}}}+4\left(-\frac{\partial f_{\nu{\bm k}}}{\partial \varepsilon_{\nu{\bm k}}}\right)\right]
    \\
    &\, \times\frac{v^{\nu}_{\mu}({\bm k})g^{\nu\bar{\nu}}_{\alpha\beta}({\bm k})+v^{\nu}_{\alpha}({\bm k})g^{\nu\bar{\nu}}_{\mu\beta}({\bm k})+v^{\nu}_{\beta}({\bm k})g^{\nu\bar{\nu}}_{\mu\alpha}({\bm k})}{\varepsilon_{\nu{\bm k}}-\varepsilon_{\bar{\nu}{\bm k}}},
    \\
    \sigma^{\mathrm{int, H}}_{\mu;\alpha\beta}=&\, \frac{e^{3}}{3V}\sum_{{\bm k}}\sum_{\nu,\bar{\nu}}\left[\frac{f_{\nu{\bm k}}-f_{\bar{\nu}{\bm k}}}{\varepsilon_{\nu{\bm k}}-\varepsilon_{\bar{\nu}{\bm k}}}
    -2\left(-\frac{\partial f_{\nu{\bm k}}}{\partial \varepsilon_{\nu{\bm k}}}\right)\right]
    \\
    &\, \times\frac{2v^{\nu}_{\mu}({\bm k})g^{\nu\bar{\nu}}_{\alpha\beta}({\bm k})-v^{\nu}_{\alpha}({\bm k})g^{\nu\bar{\nu}}_{\mu\beta}({\bm k})-v^{\nu}_{\beta}({\bm k})g^{\nu\bar{\nu}}_{\mu\alpha}({\bm k})}{\varepsilon_{\nu{\bm k}}-\varepsilon_{\bar{\nu}{\bm k}}}.
  \end{aligned}
  \label{int-O/H}
\end{equation}
We evaluate these expressions for the microscopic lattice model in Sec.~\ref{model-calculation}.

\subsection{Relation with multipoles}

To discuss the relation with the microscopic electronic degrees of freedom in the intrinsic term, we introduce the augmented multipole description~\cite{PhysRevB.98.165110, PhysRevB.98.245129}.
As the transformation properties of $\sigma^{\mathrm{O}}_{\mu;\alpha\beta}$ and $\sigma^{\mathrm{H}}_{\mu;\alpha\beta}$ are different, their corresponding multipoles are different.
Since $\sigma_{\mu;\alpha\beta}$ is the time-reversal-odd axial rank-3 tensor, the relevant multipoles are the rank-1--3 multipoles: the rank-1 magnetic toroidal dipole $(T_{x}, T_{y}, T_{z})$, the rank-2 magnetic quadrupole $(M_{u}, M_{v}, M_{yz}, M_{zx}, M_{xy})$, and the rank-3 magnetic toroidal octupole $(T^{\alpha}_{x}, T^{\alpha}_{y},T^{\alpha}_{z},T^{\beta}_{x},T^{\beta}_{y},T^{\beta}_{z}, T_{xyz})$.
The correspondence between the components of $\sigma^{\mathrm{O}}$, $\sigma^{\mathrm{H}}$ and multipole is given by 
\begin{subequations}
  \label{NLC}
  \begin{align}
    \sigma^{\mathrm{O}}=&\, \left[\begin{matrix}
    3T_{x}^{\prime}+2T_{x}^{\alpha} & T_{y}^{\prime}-T_{y}^{\alpha}-T_{y}^{\beta} & T_{z}^{\prime}-T_{z}^{\alpha}+T_{z}^{\beta} \\
    T_{x}^{\prime}-T_{x}^{\alpha}+T_{x}^{\beta} & 3T_{y}^{\prime}+2T_{y}^{\alpha} & T_{z}^{\prime}-T_{z}^{\alpha}-T_{z}^{\beta} \\
    T_{x}^{\prime}-T_{x}^{\alpha}-T_{x}^{\beta} & T_{y}^{\prime}-T_{y}^{\alpha}+T_{y}^{\beta} & 3T_{z}^{\prime}+2T_{z}^{\alpha} \\
    T_{xyz} & T_{z}^{\prime}-T_{z}^{\alpha}-T_{z}^{\beta} & T_{y}^{\prime}-T_{y}^{\alpha}+T_{y}^{\beta} \\
    T_{z}^{\prime}-T_{z}^{\alpha}+T_{z}^{\beta} & T_{xyz} & T_{x}^{\prime}-T_{x}^{\alpha}-T_{x}^{\beta} \\
    T_{y}^{\prime}-T_{y}^{\alpha}-T_{y}^{\beta} & T_{x}^{\prime}-T_{x}^{\alpha}+T_{x}^{\beta} & T_{xyz} \\
    \end{matrix}\right]^{\mathrm{T}},
    \label{NLC-O}
    \\
    \sigma^{\mathrm{H}}=&\, \left[\begin{matrix}
    0 & 2(T_{y}-M_{zx}) & 2(T_{z}+M_{xy}) \\
    2(T_{x}+M_{yz}) & 0 & 2(T_{z}-M_{xy}) \\
    2(T_{x}-M_{yz}) &  2(T_{y}+M_{zx}) & 0 \\
    M_{u}+M_{v} & -(T_{z}-M_{xy}) & -(T_{y}+M_{zx}) \\
    -(T_{z}+M_{xy}) & -M_{u}+M_{v} & -(T_{x}-M_{yz}) \\
    -(T_{y}-M_{zx}) & -(T_{x}+M_{yz}) & -2M_{v} \\
    \end{matrix}\right]^{\mathrm{T}},
    \label{NLC-H}
  \end{align}
\end{subequations}
where the matrix representation of the conductivity tensor $\sigma$ has been expressed as 
\begin{equation*}
  \sigma=
  \left[\begin{matrix}
    \sigma_{x;xx} & \sigma_{y;xx} & \sigma_{z;xx} 
    \\
    \sigma_{x;yy} & \sigma_{y;yy} & \sigma_{z;yy} 
    \\
    \sigma_{x;zz} & \sigma_{y;zz} & \sigma_{z;zz} 
    \\
    \sigma_{x;yz} & \sigma_{y;yz} & \sigma_{z;yz} 
    \\
    \sigma_{x;zx} & \sigma_{y;zx} & \sigma_{z;zx} 
    \\
    \sigma_{x;xy} & \sigma_{y;xy} & \sigma_{z;xy} 
    \\
  \end{matrix}\right]^{\mathrm{T}},
\end{equation*}
$(T_{x}, T_{y}, T_{z})$ and $(T_{x}^{\prime}, T_{y}^{\prime}, T_{z}^{\prime})$ stand for the independent magnetic toroidal dipoles and T means the transpose of a matrix.
The magnetic quadrupole (magnetic toroidal octupole) appears only in $\sigma^{\mathrm{H}}$ ($\sigma^{\mathrm{O}}$), while the magnetic toroidal dipole appears in both conductivity.
Thus, the pure INAHE is expected when the magnetic quadrupole is activated under magnetic orderings. 
We present the relation between the $\sigma_{\mu;\alpha\beta}$ and the multipoles in Table~\ref{relation-multipoles}.
\begin{table}[tbp]
  \caption{\label{relation-multipoles}
  The relation between the Ohmic/Hall part of the second-order conductivity tensor and the activated multipoles.
  M and MT multipoles represent magnetic and magnetic toroidal multipoles, respectively.
  }
  \begin{ruledtabular}
    \begin{tabular}{ccc}
      Multipole & Ohmic & Hall 
      \\
      \hline
      MT dipole & \checkmark & \checkmark 
      \\
      M quardupole & -- & \checkmark 
      \\
      MT octupole & \checkmark & -- 
      \\
    \end{tabular}
  \end{ruledtabular}
\end{table}

\section{Model Calculations}
\label{model-calculation}

As discussed in the previous section, the INAHE occurs when either a magnetic quadrupole or magnetic toroidal dipole is activated.
In this section, we evaluate the INAHE based on the microscopic lattice model to examine the key ingredients for the INAHE from the viewpoint of the electronic degrees of freedom. 
First, we construct a minimal model under the four-sublattice layered tetragonal structure, where the $\mathcal{PT}$-symmetric noncollinear AFM structures can accompany the magnetic quadrupole and magnetic toroidal dipole in Sec.~\ref{Model_Hamiltonian}.
Next, we show the numerical results for the INAHE in the model in Sec.~\ref{Model_result}.
Then, we discuss the important contributions to the INAHE of the model parameters in Sec.~\ref{Model_parameter}.

\subsection{Hamiltonian}
\label{Model_Hamiltonian}

\begin{figure}[tbp]
  \centering
  \includegraphics[width=\linewidth]{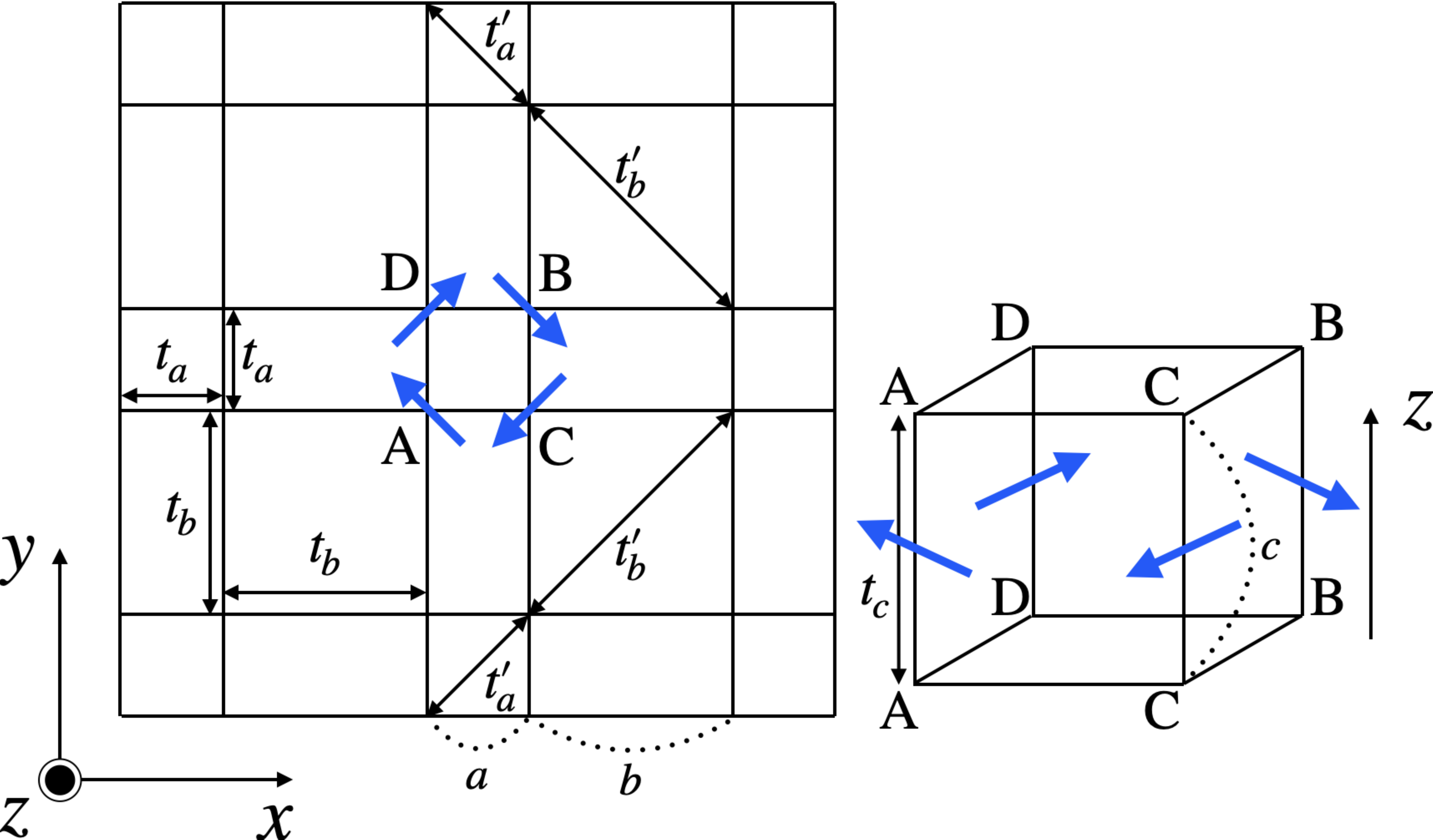}
  \caption{\label{lattice}
  The four-sublattice layered tetragonal structure with the hopping parameters $(t_{a},t_{b},t_{a}^{\prime},t_{b}^{\prime}, t_{c})$ and $g$-vector denoted as the blue arrows in our model in Eq.~(\ref{tight-binding}).
  $a$, $b$, and $c$ are the lattice constants.
  }
\end{figure}

To examine the behavior of the INAHE, we construct a minimal lattice model, as shown in Fig.~\ref{lattice}; the lattice structure consists of the four-sublattice layered tetragonal structure, whose point group belongs to $D_{\mathrm{4h}}$ ($4/mmm1'$). 
The lattice constant is set to be $a+b=2$ ($a=b=1$) and $c=1$ for simplicity.
The tight-binding Hamiltonian is given by 
\begin{equation}
  \mathcal{H}=\sum_{{\bm k}}\sum_{\gamma,\gamma^{\prime}}\sum_{\sigma,\sigma^{\prime}}\hat{c}_{{\bm k}\gamma\sigma}^{\dag}h_{\sigma\sigma^{\prime}}^{\gamma\gamma^{\prime}}({\bm k})\hat{c}^{}_{{\bm k}\gamma^{\prime}\sigma^{\prime}},
  \label{tight-binding}
\end{equation}
where $\hat{c}_{{\bm k}\gamma\sigma}^{\dag}$ and $\hat{c}_{{\bm k}\gamma\sigma}$ are the fermionic creation and annihilation operators of the wave number ${\bm k}$, the sublattice $\gamma=\mathrm{A}$--$\mathrm{D}$ and the spin $\sigma=\uparrow,\downarrow$.
The Hamiltonian matrix $[\hat{h}({\bm k})]_{(\gamma\sigma),(\gamma^{\prime}\sigma^{\prime})}\equiv h_{\sigma\sigma^{\prime}}^{\gamma\gamma^{\prime}}({\bm k})$ consists of three parts as follows: 
\begin{equation}
  \hat{h}({\bm k})=\hat{h}_{\mathrm{hop}}({\bm k})+\hat{h}_{\mathrm{ASOI}}({\bm k})+\hat{h}_{\mathrm{mf}}.
  \label{tight-binding2}
\end{equation}
The first term $\hat{h}_{\mathrm{hop}}({\bm k})$ represents the hopping term including the hoppings along the $x$ or $y$ direction ($z$ direction), $t_{a}$ and $t_{b}$ $(t_{c})$, and the hoppings along the in-plane diagonal direction, $t_{a}^{\prime}$ and $t_{b}^{\prime}$, as shown in Fig.~\ref{lattice}.
The second term $\hat{h}_{\mathrm{ASOI}}({\bm k})$ represents the sublattice-dependent ASOI, which originates from relativistic spin--orbit coupling under the lack of local inversion symmetry at each lattice site.
We here take into account the ASOI along the out-of-plane direction, which is represented by 
\begin{equation}
  \begin{aligned}
    \hat{h}_{\mathrm{ASOI}}({\bm k})=&\, \delta_{\gamma\gamma^{\prime}} {\bm g}_{\gamma} ({\bm k})\cdot ({\bm \sigma})_{\sigma\sigma^{\prime}}
    \\
    =&\, \delta_{\gamma\gamma^{\prime}}\alpha_{1}\sin{k_{z}}\left\{\hat{{\bm z}}\cdot[{\bm e}_{\gamma}\times({\bm \sigma})_{\sigma\sigma^{\prime}}] \right\},
  \end{aligned}
\end{equation}
where ${\bm e}_{\mathrm{A}}=(-1,-1,0)$, ${\bm e}_{\mathrm{B}}=(1,1,0)$, ${\bm e}_{\mathrm{C}}=(1,-1,0)$, ${\bm e}_{\mathrm{D}}=(-1,1,0)$, $\hat{\bm z}=(0,0,1)$, and ${\bm \sigma}=(\sigma_{x},\sigma_{y},\sigma_{z})$ is the vector of the Pauli matrix in spin space.
${\bm g}_{\gamma}({\bm k})$ is the so-called (sublattice-dependent) $g$-vector, and its direction in each sublattice is presented in Fig.~\ref{lattice}, which forms the vortex structure to satisfy fourfold rotational symmetry~\cite{PhysRevB.90.024432, PhysRevB.104.134420, PhysRevB.103.054416, JPSJ.91.123701}.
It is noted that $\sum_{\gamma}{\bm g}_{\gamma}({\bm k})=0$ owing to the presence of global inversion symmetry.

\begin{figure}[tbp]
  \includegraphics[width=\linewidth]{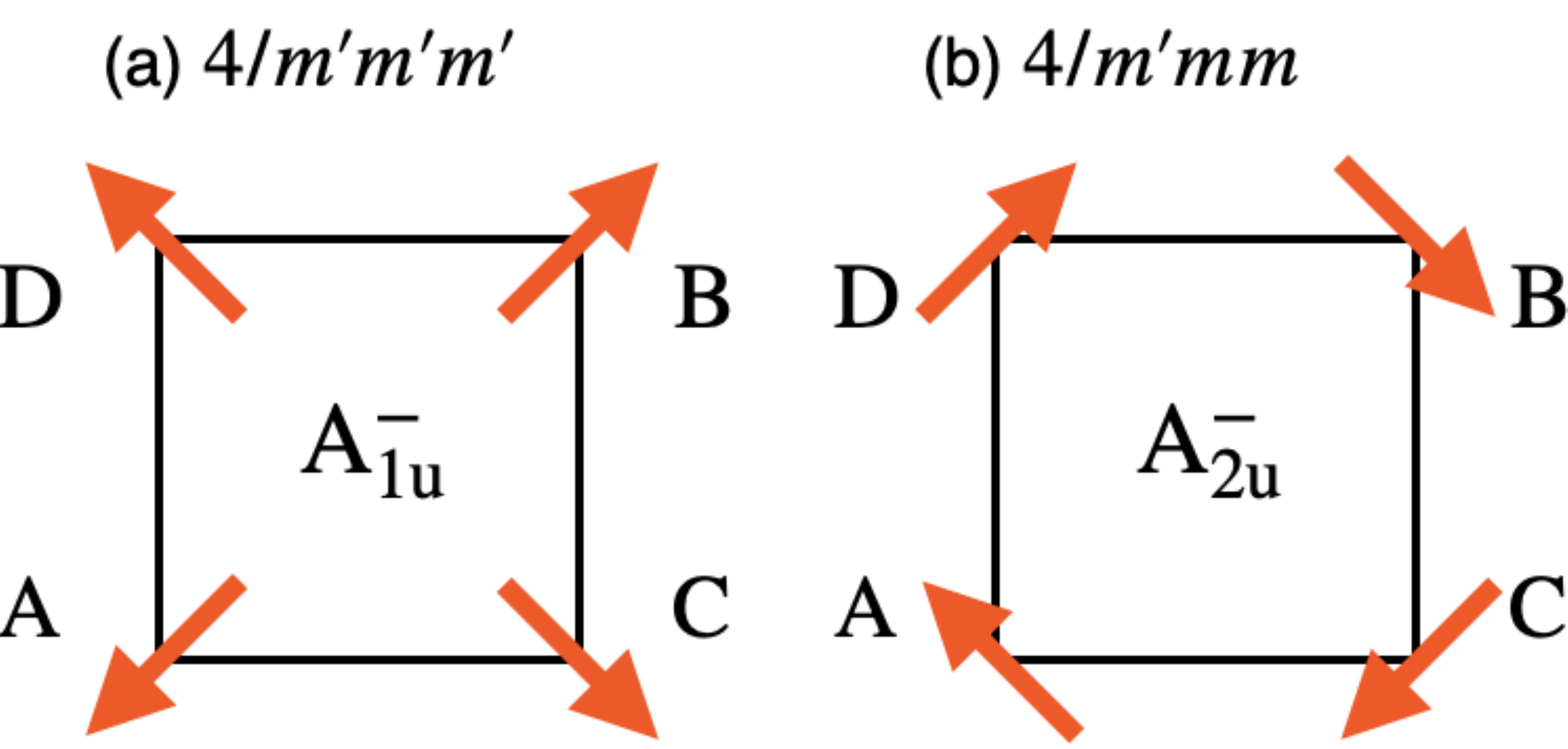}
  \caption{\label{order}The noncollinear spin configurations to accompany (a) the magnetic quadrupole and (b) the magnetic toroidal dipole.
  The red arrows represent the spin moments.
  We also show the magnetic point groups and irreducible representations in $D_{\mathrm{4h}}$.}
\end{figure}

The third term $\hat{h}_{\mathrm{mf}}$ stands for the molecular field (MF) corresponding to the magnetic order, which arises from the MF approximation to the Coulomb interaction.
We consider two types of noncollinear magnetic textures with magnetic quadrupole $M_{u}$ and magnetic toroidal dipole $T_{z}$ as shown in Figs.~\ref{order}(a) and \ref{order}(b), respectively.
The expression of the MF Hamiltonian matrix is given by 
\begin{equation}
  \hat{h}_{\mathrm{mf}}=
  \begin{cases}
    h_{\mathrm{AF}}\delta_{\gamma\gamma^{\prime}}{\bm e}_{\gamma}\cdot({\bm \sigma})_{\sigma\sigma^{\prime}}
    \\
    h_{\mathrm{AF}}\delta_{\gamma\gamma^{\prime}}\hat{\bm z}\cdot[{\bm e}_{\gamma}\times({\bm \sigma})_{\sigma\sigma^{\prime}}]
    \\
  \end{cases}
  \label{molecular_field}
\end{equation}
with the magnitude of the MF $h_{\mathrm{AF}}$.
The first row corresponds to the MF for the magnetic quadrupole $M_{u}$, while the second row corresponds to that for the magnetic toroidal dipole $T_{z}$.
In the case of $M_{u}$ $(T_{z})$, the MPG reduces to $4/m^{\prime}m^{\prime}m^{\prime}$ ($4/m^{\prime}mm$)~\cite{PhysRevB.104.054412}; $\mathcal{P}$ symmetry is broken while keeping $\mathcal{PT}$ symmetry in both cases.
These odd-parity multipoles have been recently discussed in AFM metals since they give rise to unconventional off-diagonal responses and quantum transports, such as the magnetoelectric effect~\cite{PhysRevB.96.064432,thole2018magnetoelectric, JPSJ.87.033702, JPSJ.89.033703, PhysRevB.104.045117}, nonlinear Hall effect~\cite{arxiv.2205.05555}, and nonreciprocal spin transport~\cite{PhysRevB.106.024405}.

For later discussion in Sec.~\ref{Model_parameter}, we introduce the multipole description for the Hamiltonian in Eq.~(\ref{tight-binding2}).
By using the symmetry-adapted multipole basis~\cite{JPSJ.88.123702, PhysRevB.101.220403, PhysRevB.102.144441}, the Hamiltonian can be expressed as the product form of the cluster and bond degrees of freedom (denoted as cluster multipole and bond multipole, respectively) and the spin degree of freedom.
For example, the onsite degrees of freedom in the Hamiltonian matrix is described by the cluster electric multipole $Q^{(\mathrm{c})}$, while the off-site degrees of freedom are described by the bond electric multipole $Q^{(\mathrm{b}n)}$ and bond magnetic toroidal multipole $T^{(\mathrm{b}n)}$, where the superscript $n$ represents the index for the bond.
The matrix representation of the relevant multipoles is given in Appendix~\ref{matrix_element}. 

Then, the hopping Hamiltonian $\hat{h}_{\mathrm{hop}}({\bm k})$ can be expressed as follows: 
\begin{equation}
  \begin{aligned}
    \hat{h}_{\mathrm{hop}}({\bm k})=&\, \hat{h}_{1}({\bm k})+\hat{h}_{2}({\bm k})+\hat{h}_{z}
    ({\bm k}),
    \\
    \hat{h}_{1}({\bm k})=&\, (t_{a}+t_{b})(c_{x}+c_{y})Q_{0}^{(\mathrm{b}1)}-(t_{a}+t_{b})(c_{x}-c_{y})Q_{v}^{(\mathrm{b}1)}
    \\
    &\, +(t_{a}-t_{b})s_{x}T_{x}^{(\mathrm{b}1)}+(t_{a}-t_{b})s_{y}T_{y}^{(\mathrm{b}1)},
    \\
    \hat{h}_{2}({\bm k})=&\, (t_{a}^{\prime}+t_{b}^{\prime})c_{x}c_{y}Q_{0}^{(\mathrm{b}2)}-(t_{a}^{\prime}+t_{b}^{\prime})s_{x}s_{y}Q_{xy}^{(\mathrm{b}2)}
    \\
    &\, +(t_{a}^{\prime}-t_{b}^{\prime})s_{x}c_{y}T_{x}^{(\mathrm{b}2)}+(t_{a}^{\prime}-t_{b}^{\prime})c_{x}s_{y}T_{y}^{(\mathrm{b}2)},
    \\
    \hat{h}_{z}({\bm k})=&\, 2t_{c}c_{z}Q_{0}^{(\mathrm{c})},
    \label{hopping_multipole}
  \end{aligned}
\end{equation}
where the abbreviation as $s_{\alpha}=\sin{k_{\alpha}},c_{\alpha}=\cos{k_{\alpha}}$ for $\alpha=x,y,z$ is used for notational simplicity.
The superscript (c) in the multipole matrices represents the cluster multipole, while (b1) and (b2) represent the bond multipole in $\hat{h}_{1}(\bm{k})$ and $\hat{h}_{2}(\bm{k})$, respectively.
Similarly, the ASOI and MF terms are expressed as 
\begin{equation}
  \hat{h}_{\mathrm{ASOI}}({\bm k})
  =\alpha_{1}s_{z}(Q_{y}^{(\mathrm{c})}\sigma_{x}-Q_{x}^{(\mathrm{c})}\sigma_{y})
  \equiv\alpha_{1}s_{z}T_{z}^{(\mathrm{c})},
  \label{asoi_multipole}
\end{equation}
and
\begin{equation}
  \hat{h}_{\mathrm{mf}}=
  \begin{cases}
    h_{\mathrm{AF}}(Q_{x}^{(\mathrm{c})}\sigma_{x}+Q_{y}^{(\mathrm{c})}\sigma_{y})\equiv h_{\mathrm{AF}}M_{u}^{(\mathrm{c})}
    \\
    h_{\mathrm{AF}}(Q_{y}^{(\mathrm{c})}\sigma_{x}-Q_{x}^{(\mathrm{c})}\sigma_{y})\equiv h_{\mathrm{AF}}T_{z}^{(\mathrm{c})}.
    \\
  \end{cases}
  \label{mf_multipole}
\end{equation}
We summarize the correspondence between multipoles included in the Hamiltonian and their irreducible representations in Table~\ref{irrep}.
It is noted that the multipole degree of freedom belonging to the $\mathrm{B}_{1g}^{+}$ ($\mathrm{B}_{2g}^{+}$) representation appears only in $\hat{h}_{{1}}({\bm k})$ [$\hat{h}_{{2}}({\bm k})$].

\begin{table}[htbp]
  \caption{\label{irrep}The correspondence between multipoles included in the Hamiltonian and irreducible representations (IRs) in $D_{\mathrm{4h}}$. 
  The superscript $\pm$ of the irreducible representation denotes the parity with respect to $\mathcal{T}$ operation.}
  \begin{ruledtabular}
    \begin{tabular}{lll}
      Hamiltonian & Multipole & IR 
      \\
      \hline
      $\hat{h}_{1}({\bm k})$ & $Q_{0}^{(\mathrm{b}1)}$ & $\mathrm{A}_{1g}^{+}$
      \\
      & $Q_{v}^{(\mathrm{b}1)}$ & $\mathrm{B}_{1g}^{+}$
      \\
      & $T_{x}^{(\mathrm{b}1)}, T_{y}^{(\mathrm{b}1)}$ & $\mathrm{E}_{u}^{-}$
      \\
      \hline
      $\hat{h}_{2}({\bm k})$ & $Q_{0}^{(\mathrm{b}2)}$ & $\mathrm{A}_{1g}^{+}$
      \\
      & $Q_{xy}^{(\mathrm{b}2)}$ & $\mathrm{B}_{2g}^{+}$
      \\
      & $T_{x}^{(\mathrm{b}2)}, T_{y}^{(\mathrm{b}2)}$ & $\mathrm{E}_{u}^{-}$
      \\
      \hline
      $\hat{h}_{z}({\bm k})$ & $Q_{0}^{(\mathrm{c})}$ & $\mathrm{A}_{1g}^{+}$
      \\
      \hline
      $\hat{h}_{\mathrm{ASOI}}({\bm k})$ & $T_{z}^{(\mathrm{c})}$ & $\mathrm{A}_{2u}^{-}$
      \\
      \hline
      $\hat{h}_{\mathrm{mf}}$ & $M_{u}^{(\mathrm{c})}$ & $\mathrm{A}_{1u}^{-}$
      \\
      & $T_{z}^{(\mathrm{c})}$ & $\mathrm{A}_{2u}^{-}$
      \\
    \end{tabular}
  \end{ruledtabular}
\end{table}

\subsection{Numerical Results}
\label{Model_result}

\begin{figure}[htbp]
  \includegraphics[width=\linewidth]{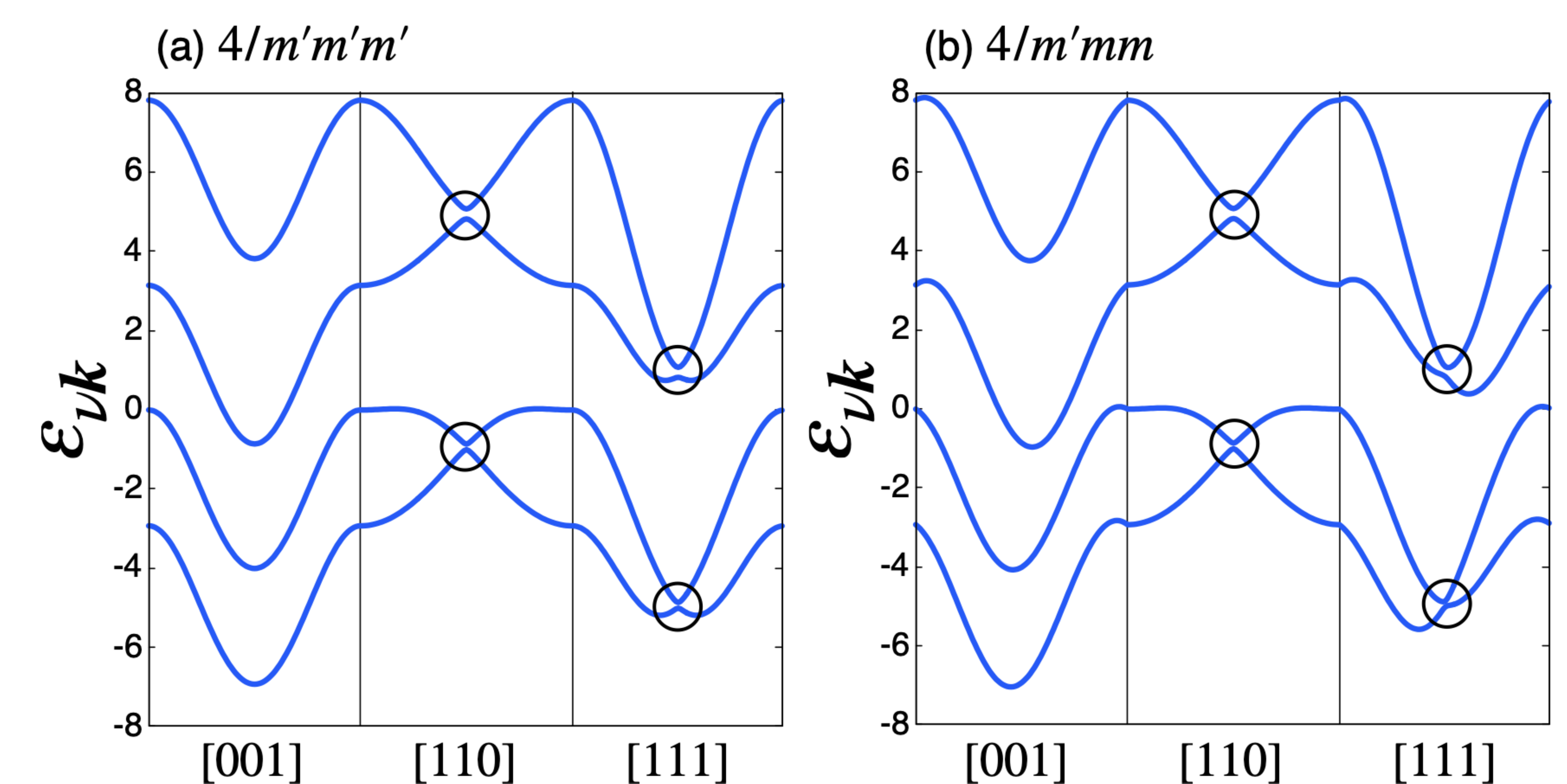}
  \caption{\label{band_structure}
  Band structures for the magnetic orderings with (a) $M_{u}$ under the magnetic point group (MPG) $4/m^{\prime}m^{\prime}m^{\prime}$ and (b) $T_{z}$ under the MPG $4/m^{\prime}mm$.
  In each panel, the wavenumber $(k_{x},k_{y},k_{z})$ dependence is as follows: $(k_{x},k_{y},k_{z})=(0,0,2s)$ in [001], $(s,s,0)$ in [110], and $(s,s,2s)$ in [111] with $0\leq s\leq \pi$.
  The circles indicate the narrow band gap region.
  }
\end{figure}

First, we show the band structure under the noncollinear AFM orderings for $h_{\mathrm{AF}} \neq 0$ in Figs.~\ref{band_structure}(a) and \ref{band_structure}(b), where the model parameters are set as
\begin{equation}
  \begin{aligned}
    &\, t_{a}=1,\ 
    t_{b}=0.9,\ 
    t_{a}^{\prime}=0.5,\ 
    t_{b}^{\prime}=0.3,\ 
    t_{c}=1,
    \\
    &\, \alpha_{1}=0.5,\ 
    h_{\mathrm{AF}}=2.
  \end{aligned}
  \label{parameters}
\end{equation}
There are two characteristic features in the band structures:
One is the asymmetric band modulation along the $k_z$ line with respect to the origin $(k_x, k_y, k_z)=(0,0,0)$ for the AFM state with $T_{z}$, as shown in Fig.~\ref{band_structure}(b). 
Meanwhile, the band structure is symmetric in the case of the AFM state with $M_{u}$, as shown in Fig.~\ref{band_structure}(a).
The appearance of the asymmetric band structure in the AFM state with $T_{z}$ is owing to a parallel alignment of the spin moment and the $g$-vector at each sublattice, which leads to an effective coupling~\cite{JPSJ.83.014703, PhysRevB.90.024432, JPSJ.84.064717}.
The other is the line-node Dirac-type band dispersion with a narrow gap along the $k_{z}$ line at $(k_{x},k_{y})=(\pi/2,\pi/2)$, as denoted by the circles in Figs.~\ref{band_structure}(a) and \ref{band_structure}(b).
This is attributed to the fact that the noncollinear spin textures in Figs.~\ref{order}(a) and \ref{order}(b) are related to the $\pi$-flux state in the case of $t_{a}=t_{b}$ and $t_{a}^{\prime}=t_{b}^{\prime}$~\cite{PhysRevB.62.13816, PhysRevB.91.075104}; the narrow gap structure is owing to the breathing property of the lattice structure.

Next, we numerically evaluate the intrinsic term in Eq.~(\ref{int-O/H}) to investigate the INAHE.
We set $e=\hbar=1$, the temperature $T=0.01$ and choose the ${\bm k}$ mesh $N=2^{2}V/(a+b)^{2}c$ as $175^{2}\times 1000$.
In the AFM state with $M_{u}$, $\sigma^{\mathrm{H}}_{x;yz}=-\sigma^{\mathrm{H}}_{y;zx}$ becomes nonzero from Eq.~(\ref{NLC}) since only $M_{u}$ is activated among the relevant multipoles under $4/m^{\prime}m^{\prime}m^{\prime}$.
Meanwhile, $\sigma^{\mathrm{O}}$ vanishes owing to the absence of the magnetic toroidal dipole and octupole.
Thus, the pure INAHE is expected in this case.

\begin{figure}[htbp]
  \includegraphics[width=\linewidth]{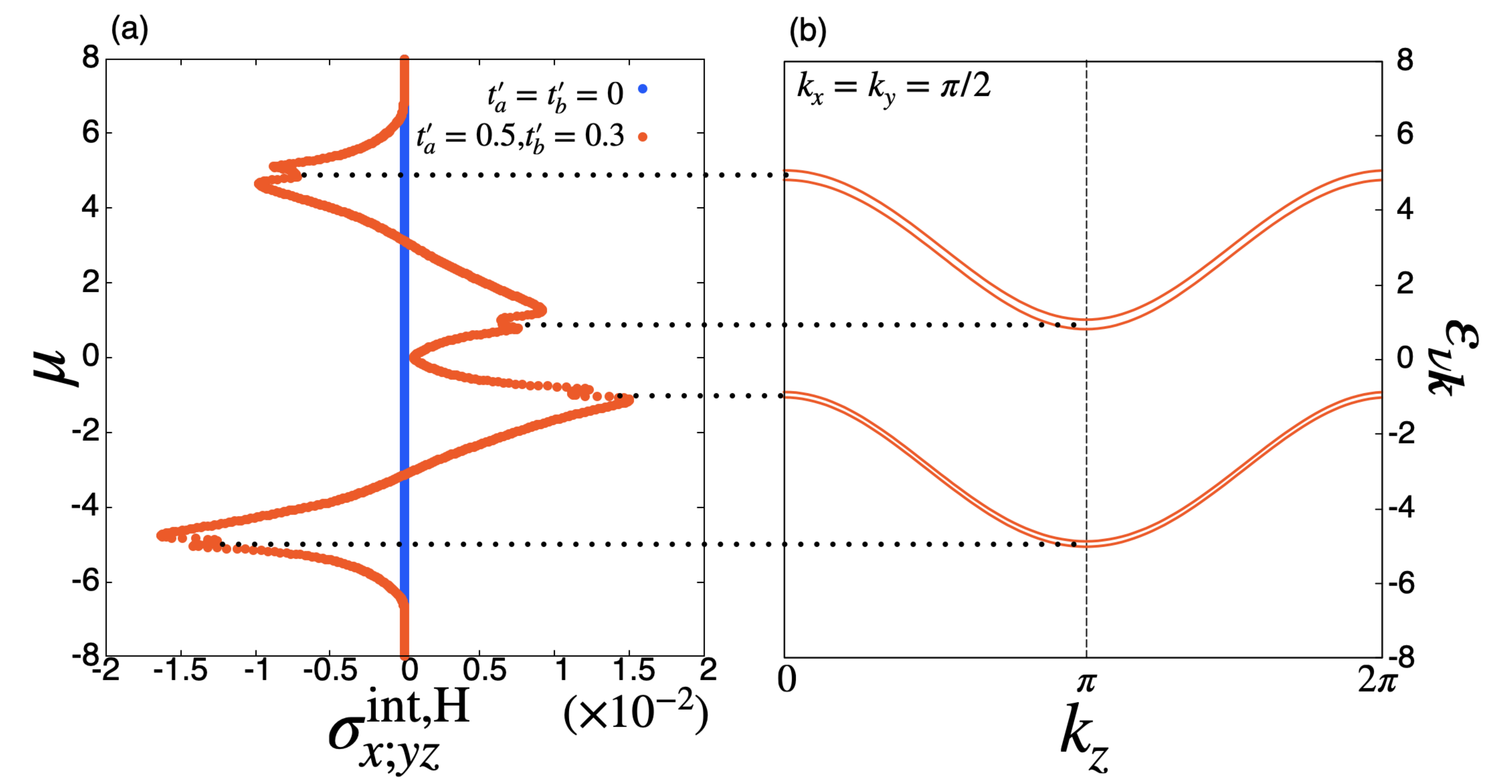}
  \caption{\label{Mu_result}
  (a) Chemical potential $\mu$ dependence of $\sigma_{x;yz}^{\mathrm{int, H}}$ evaluated by Eq.~(\ref{int-O/H}) with (red) and without (blue) the diagonal hoppings $t_{a}^{\prime}$and $t_{b}^{\prime}$ at $T=0.01$.
  (b) Band structure along with the $k_{x}=k_{y}=\pi/2$ line.
  The black dotted lines represent the correspondence between the region of the narrow band gap at the band edges and the conductivity peak structures.
  }
\end{figure}

The red dots in Fig.~\ref{Mu_result}(a) show the chemical potential $\mu$ dependence of $\sigma_{x;yz}^{\mathrm{int, H}}$ by using the model parameters in Eq.~(\ref{parameters}).
As expected from the MPG symmetry in the presence of $M_{u}$, the system exhibits nonzero $\sigma_{x;yz}^{\mathrm{int, H}}$ irrespective of $\mu$; there are roughly four peak structures at $\mu \simeq -5, -1, +1$, and $+5$.
Since the band structure in this case is symmetric, the microscopic origin of $\sigma_{x;yz}^{\mathrm{int, H}}$ is different from that of the Drude-type conductivity $\sigma^{\mathrm{D}}$, which arises from the asymmetric band modulation.
We find that these peak structures correspond well to the narrow gap region at the band edges, as shown in Fig.~\ref{Mu_result}(b), where the black dotted lines represent their correspondence.
Such a correspondence is reasonable in terms of the expression in Eq.~(\ref{int-O/H}); the narrow gap leading to a small denominator yields a large $\sigma_{x;yz}^{\mathrm{int, H}}$.
We find that the dominant contribution arises from the Dirac-type band dispersion in the $k_{x}$--$k_{y}$ plane in Fig.~\ref{Mu_result}(b), which means that the QM dipole $v^{\nu}_{x}({\bm k})g^{\nu\bar{\nu}}_{yz}({\bm k})=-v^{\nu}_{y}({\bm k}^{\prime})g^{\nu\bar{\nu}}_{zx}({\bm k}^{\prime})$ with ${\bm k}^{\prime}\equiv (-k_{y},k_{x},k_{z})$ becomes important for $\sigma_{x;yz}^{\mathrm{int, H}}$.
This behavior coincides with the semiclassical analysis~\cite{PhysRevLett.127.277201, PhysRevLett.127.277202}.
In addition, the contributions of the Fermi sea and Fermi surface terms are comparable to each other, as discussed in Appendix~\ref{sea-surface}.
We also discuss the results for the case of a large MF in Appendix~\ref{sea-surface}, where we confirm that the main contribution comes from near the Dirac point as well in this case.
When we turn off the ASOI, the QM dipole $v_{x}^{\nu}({\bm k})g_{yz}^{\nu\bar{\nu}}({\bm k})$ completely vanishes, which indicates that the ASOI is important for the INAHE.

The above situation qualitatively changes when different hopping parameters are set.
For example, $\sigma_{x;yz}^{\mathrm{int, H}}$ vanishes by setting $t_{a}^{\prime}=t_{b}^{\prime}=0$ while the other parameters remain the same, as shown by the blue dots in Fig.~\ref{Mu_result}(a).
This result indicates that the diagonal hopping is necessary to induce $\sigma_{x;yz}^{\mathrm{int, H}}$, which is not accounted for by the symmetry argument and the geometry of the band structure.
We will discuss this point in Sec.~\ref{Model_parameter}.

Let us discuss the other AFM state with $T_{z}$ in Fig.~\ref{order}(b), where the MPG reduces to $4/m^{\prime}mm$.
Since the relevant active multipoles are $T_{z}$ and $T_z^\alpha$ in this case, there are three independent tensor components in $\sigma_{\mu;\alpha\beta}$: $\sigma^{\mathrm{O}}_{z;zz}$, $\sigma^{\mathrm{O}}_{z;xx}=\sigma^{\mathrm{O}}_{z;yy}=\sigma^{\mathrm{O}}_{x;zx}=\sigma^{\mathrm{O}}_{y;yz}$, and $\sigma^{\mathrm{H}}_{z;xx}=\sigma^{\mathrm{H}}_{z;yy}=-2\sigma^{\mathrm{H}}_{x;zx}=-2\sigma^{\mathrm{H}}_{y;yz}$ in Eq.~(\ref{NLC}).
In other words, not only $\sigma^{\mathrm{H}}$ but also $\sigma^{\mathrm{O}}$ are expected to be nonzero in contrast to the $M_{u}$ case.
We ignore the contribution from the Drude term in order to focus on the behavior of $\sigma^{\mathrm{int}}_{\mu;\alpha\beta}$~\cite{PhysRevResearch.2.043081, PhysRevB.105.155157}, although it can also be finite owing to the asymmetric band structure in Fig.~\ref{band_structure}(b).

\begin{figure*}[htbp]
  \includegraphics[width=\linewidth]{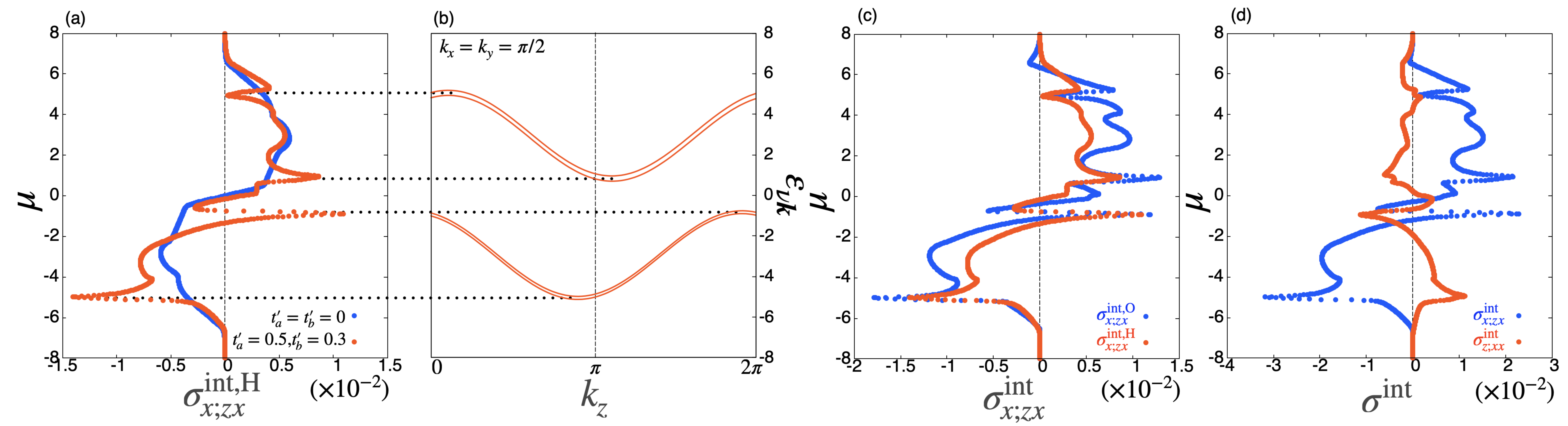}
  \caption{\label{Tz_result}
  (a) Chemical potential $\mu$ dependence of $\sigma_{x;zx}^{\mathrm{int, H}}$ evaluated by Eq.~(\ref{int-O/H}) with (red) and without (blue) the diagonal hoppings $t_{a}^{\prime}$ and $t_{b}^{\prime}$ at $T=0.01$.
  (b) Band structure along with the $k_{x}=k_{y}=\pi/2$ line.
  The black dotted lines represent the correspondence between the region of the narrow band gap at the band edges and the conductivity peak structures.
  (c) $\mu$ dependence of the Ohmic (blue) and Hall (red) part in $\sigma_{x;zx}^{\mathrm{int}}$.
  (d) Results for the total intrinsic conductivity $\sigma_{x;zx}^{\mathrm{int}}=\sigma_{x;zx}^{\mathrm{int, O}}+\sigma_{x;zx}^{\mathrm{int, H}}$ (blue) and $\sigma_{z;xx}^{\mathrm{int}}=\sigma_{z;xx}^{\mathrm{int, O}}+\sigma_{z;xx}^{\mathrm{int, H}}=\sigma_{x;zx}^{\mathrm{int, O}}-2\sigma_{x;zx}^{\mathrm{int, H}}$(red).}
\end{figure*}

Figure \ref{Tz_result}(a) shows the $\mu$ dependence of $\sigma_{x;zx}^{\mathrm{int, H}}$ under the AFM with $T_{z}$, where the same hopping and ASOI parameters are used as those with $M_{u}$.
Similar to the case with $M_{u}$, $\sigma_{x;zx}^{\mathrm{int, H}}$ becomes nonzero, which exhibits the peak or dip structures at the band edges with a narrow gap, as shown by the dotted black lines in Figs.~\ref{Tz_result}(a) and \ref{Tz_result}(b).
Thus, the narrow gap structure in the band dispersion plays an important role in enhancing $\sigma_{x;zx}^{\mathrm{int, H}}$.
The complicated $\mu$ dependence of $\sigma_{x;zx}^{\mathrm{int ,H}}$ for $-4\lesssim \mu\lesssim 4$ compared to $\sigma_{x;yz}^{\mathrm{int, H}}$ in the case of $M_{u}$ is attributed to the contributions between the middle two bands.
Indeed, similar $\mu$ dependence to the $M_u$ case is obtained for a large MF, as shown in Appendix~\ref{sea-surface}.
It is noted that $\sigma_{x;zx}^{\mathrm{int ,H}}$ vanishes in the absence of the ASOI, as $v_{z}^{\nu}({\bm k})g_{xx}^{\nu\bar{\nu}}({\bm k})$ cancels out with $v_{z}^{\bar{\nu}}({\bm k})g_{xx}^{\bar{\nu}\nu}({\bm k})$. 

On the other hand, there are two qualitative differences from the result with $M_{u}$. 
One is the model parameter dependence; $\sigma_{x;zx}^{\mathrm{int, H}}$ remains nonzero values even for $t_{a}^{\prime}=t_{b}^{\prime}=0$, as shown in Fig.~\ref{Tz_result}(a). 
This indicates the difference in the microscopic origin of the INAHE for the magnetic quadrupole and magnetic toroidal dipole, which will be discussed in Sec.~\ref{Model_parameter}.

Another is the additional contribution to $\sigma_{x;zx}^{\mathrm{int}}$ from the Ohmic part $\sigma_{x;zx}^{\mathrm{int, O}}$ in measurements.
Figure~\ref{Tz_result}(c) shows the $\mu$ dependence of $\sigma_{x;zx}^{\mathrm{int, O}}$ with the result for $\sigma_{x;zx}^{\mathrm{int, H}}$ in Fig.~\ref{Tz_result}(a) for reference.
Compared to both results, one finds that $\sigma_{x;zx}^{\mathrm{int, O}}$ has a similar $\mu$ dependence as $\sigma_{x;zx}^{\mathrm{int, H}}$ so as to enhance the total contribution $\sigma_{x;zx}^{\mathrm{int}}=\sigma_{x;zx}^{\mathrm{int, O}}+\sigma_{x;zx}^{\mathrm{int, H}}$, as shown by the blue dots in Fig.~\ref{Tz_result}(d).
On the other hand, it is noted that there can be a cancellation depending on the tensor components.
For example, the result for $\sigma_{z; xx}^{\mathrm{int}}=\sigma_{z; xx}^{\mathrm{int, O}}+\sigma_{z; xx}^{\mathrm{int, H}}=\sigma_{x;zx}^{\mathrm{int, O}}-2\sigma_{x;zx}^{\mathrm{int, H}}$ is presented by the red dots in Fig.~\ref{Tz_result}(d), where the absolute values tend to be smaller than those in $\sigma_{x;zx}^{\mathrm{int}}$.

\subsection{Effective closed path}
\label{Model_parameter}

In the previous section, the numerical result indicates the characteristic hopping parameter dependence in the intrinsic term in $\sigma_{\mu;\alpha\beta}$ for the AFM state with $M_{u}$.
In this section, we elucidate the essential model parameters for the INAHE in both AFM cases.
For that purpose, we evaluate the following quantity:
\begin{equation}
  \Gamma^{ijk}_{\mu;\alpha,\beta}=\sum_{{\bm k}}
  \mathrm{Tr}{\left[
    \hat{v}_{\mu}({\bm k})\hat{h}^{i}({\bm k})\hat{v}_{\alpha}({\bm k})\hat{h}^{j}({\bm k})\hat{v}_{\beta}({\bm k})\hat{h}^{k}({\bm k})
  \right]},
  \label{essential}
\end{equation}
which was obtained by the expansion of the nonlinear conductivity tensor~\cite{JPSJ.91.014701}; the order of the expansion is $i+j+k$, where $i,j,k$ are integers.
The real part of $\Gamma^{ijk}_{\mu;\alpha,\beta}$ contributes to the INAHE.

Let us consider the conditions for nonzero $\Gamma^{ijk}_{\mu;\alpha,\beta}$ in Eq.~(\ref{essential}).
As the INAHE is induced by the onset of the AFM ordering, $\Gamma^{ijk}_{\mu;\alpha,\beta}$ must include the contribution of $h_{\mathrm{AF}}$; otherwise, $\Gamma^{ijk}_{\mu;\alpha,\beta}=0$.
In addition, its $h_{\mathrm{AF}}$ dependence should be an odd function since the AFM state with the different domain leads to the opposite sign of $\sigma^{\mathrm{int}}_{\mu;\alpha\beta}$.
In this situation, the trace in Eq.~(\ref{essential}) should include other spin-dependent terms to make the trace, at least in spin space, nonzero.
In the present model in Eq.~(\ref{tight-binding2}), such a term corresponds to the ASOI in Eq.~(\ref{asoi_multipole}).
The remaining condition for nonzero $\Gamma^{ijk}_{\mu;\alpha,\beta}$ depends on the hopping elements in Eq.~(\ref{hopping_multipole}) so that the trace in sublattice space becomes nonzero.

From the symmetry viewpoint, the trace can become nonzero when the irreducible representation of the direct product in the trace belongs to the totally symmetric representation, i.e., $\mathrm{A}_{1g}^{+}$.
The direct product representation of the ASOI and MF terms is given by $\mathrm{A}_{2u}^{-}\otimes\mathrm{A}_{1u}^{-}=\mathrm{A}_{2g}^{+}$ in the case of the AFM order with $M_{u}$ and $\mathrm{A}_{2u}^{-}\otimes\mathrm{A}_{2u}^{-}=\mathrm{A}_{1g}^{+}$ in the case of that with $T_{z}$ (see Table~\ref{irrep}).
In order to construct the $\mathrm{A}_{2g}^{+}$ representation in the former, the irreducible representation consisting of the hoppings should include both $\mathrm{B}_{1g}^{+}$ and $\mathrm{B}_{2g}^{+}$.
Thus, one notices that the $Q_{xy}^{(\mathrm{b}2)}$-type hopping corresponding to $\mathrm{B}_{2g}^{+}$ in $\hat{h}_{2}$ is essential to construct the $\mathrm{A}_{1g}^{+}$ representation for the AFM state with $M_{u}$.
In this way, the diagonal hopping is necessary to obtain nonzero $\sigma_{\mu;\alpha\beta}^{\mathrm{int, H}}$ in the AFM state with $M_{u}$, which is consistent with the numerical results in Fig.~\ref{Mu_result}(a).
Meanwhile, there is no such a constraint for the AFM state with $T_{z}$, as it needs just the $\mathrm{A}_{1g}^{+}$, which is usually included in any hoppings.

Based on the above symmetry analysis, we directly evaluate $\Gamma^{ijk}_{x;y,z}$ to find the important model parameters for $\sigma_{x;yz}^{\mathrm{int, H}}$.
In the case of the AFM state with $M_{u}$, the lowest-order contribution to $\sigma_{x;yz}^{\mathrm{int, H}}$ arises at $(i,j,k)=(0,2,1)$, where $\Gamma^{021}_{x;y,z}$ is given by  
\begin{equation}
  \begin{aligned}
    \Gamma^{021}_{x;y,z}=&\, \sum_{{\bm k}}
    \mathrm{Tr}{\left[
      \hat{v}_{x}({\bm k})
      \hat{v}_{y}({\bm k})\hat{h}^{2}({\bm k})\hat{v}_{z}({\bm k})\hat{h}({\bm k})
    \right]}
    \\
    =&\, 32h_{\mathrm{AF}}\alpha_{1}t_{c}(t_{a}^{2}t_{a}^{\prime}+t_{b}^{2}t_{b}^{\prime}).
  \end{aligned}
  \label{low-Mu}
\end{equation}
This expression indicates that nonzero $ \sigma_{x;yz}^{\mathrm{int, H}}$ is obtained when $h_{\mathrm{AF}} \neq 0$, $\alpha_1 \neq 0$, and $t_{a}^{\prime},t_{b}^{\prime} \neq 0$, as expected in the above symmetry argument.
Moreover, one finds that $t_{c} \neq 0$ is also important to cause nonzero $ \sigma^{\mathrm{int, H}}_{x;yz}$.  
Such a relation holds for higher-order contributions of $\Gamma^{ijk}_{x;y,z}$.

\begin{figure}[htbp]
  \includegraphics[width=\linewidth]{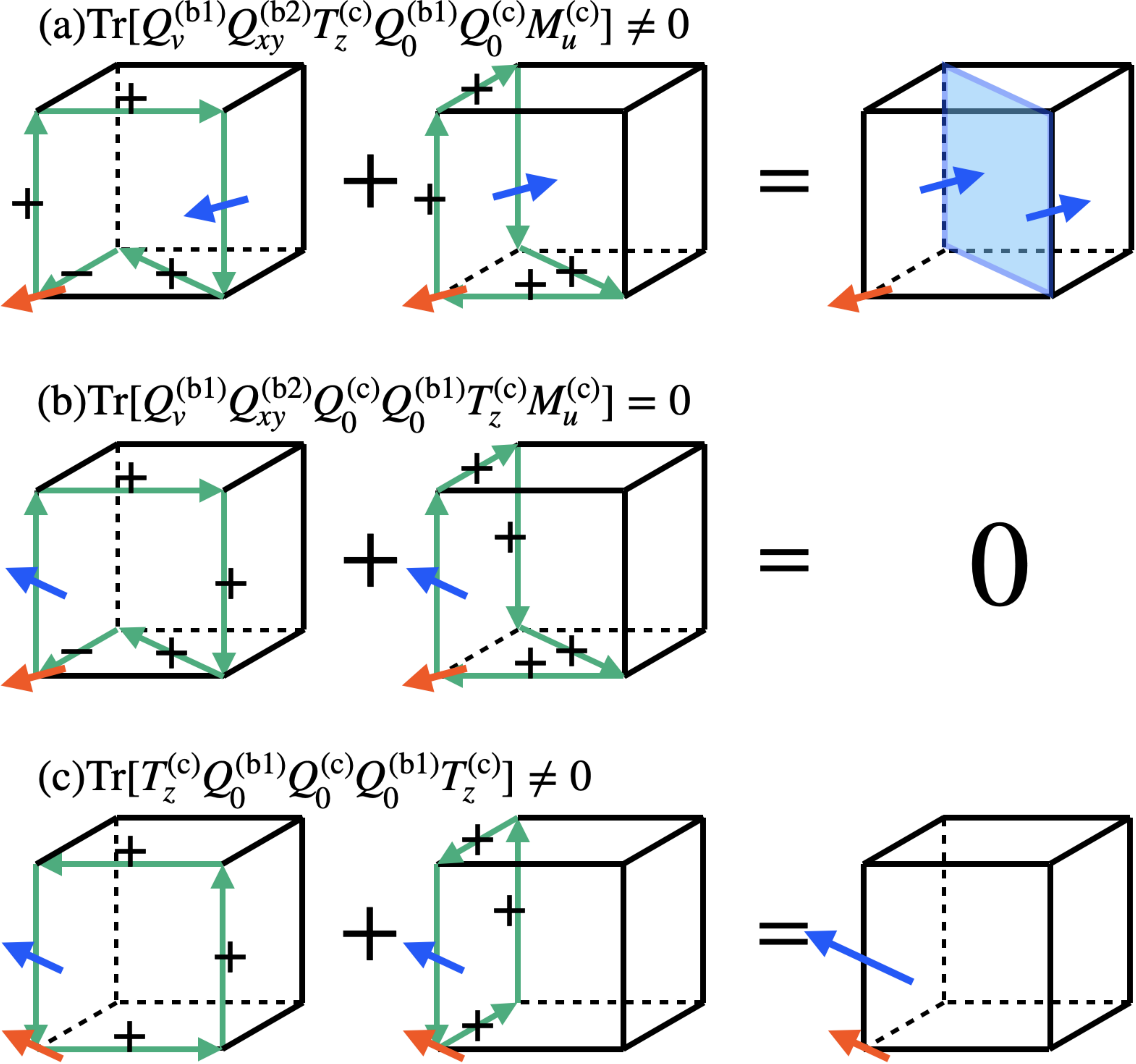}
  \caption{\label{path}
  The closed paths contributing to the INAHE. 
  As a representative, we show the paths starting from sublattice A.
  (a), (b), and (c) correspond to Eqs. (\ref{path-Mu}), (\ref{path-Mu-zero}), and (\ref{path-Tz}), respectively.
  The red (blue) arrows represent the spin moments ($g$-vectors).
  The signs on the bonds indicate the phase of each hopping.
  }
\end{figure}

To further understand the microscopic process contributing to $ \sigma_{x;yz}^{\mathrm{int, H}}$, we investigate the important hopping paths in real space.
By analyzing the expression in Eq.~(\ref{low-Mu}), one finds one of the contributing closed paths, which is given by 
\begin{equation}
  \mathrm{Tr}[Q_{v}^{(\mathrm{b}1)}Q_{xy}^{(\mathrm{b}2)}T_{z}^{(\mathrm{c})}Q_{0}^{(\mathrm{b}1)}Q_{0}^{(\mathrm{c})}M_{u}^{(\mathrm{c})}]\neq 0,
  \label{path-Mu}
\end{equation}
where there is a following correspondence: $\hat{v}_{x}({\bm k}) \leftrightarrow Q_{v}^{(\mathrm{b}1)}$, $\hat{v}_{y}({\bm k}) \leftrightarrow Q_{xy}^{(\mathrm{b}2)}$, $\hat{h}^{2}({\bm k}) \leftrightarrow T_{z}^{(\mathrm{c})}Q_{0}^{(\mathrm{b}1)}$, $\hat{v}_{z}({\bm k}) \leftrightarrow Q_{0}^{(\mathrm{c})}$, and $\hat{h}({\bm k})\leftrightarrow M_{u}^{(\mathrm{c})}$.
We pictorially plot this process in Fig.~\ref{path}(a) starting at the sublattice $\mathrm{A}$.
It is noted that an order of the multipole matrices is important; for example, a different sequence like 
\begin{equation}
  \mathrm{Tr}[Q_{v}^{(\mathrm{b}1)}Q_{xy}^{(\mathrm{b}2)}Q_{0}^{(\mathrm{c})}Q_{0}^{(\mathrm{b}1)}T_{z}^{(\mathrm{c})}M_{u}^{(\mathrm{c})}]
  \label{path-Mu-zero}
\end{equation}
gives no contribution to $ \sigma_{x;yz}^{\mathrm{int, H}}$.
The closed path in this case is shown in Fig.~\ref{path}(b).
By closely looking at the difference between Eqs. (\ref{path-Mu}) and (\ref{path-Mu-zero}), one finds that the relation of the spin and the $g$-vector is different from each other.
In the former case, the spin moment and $g$-vector are coupled in parallel (or antiparallel), which gives a totally symmetric representation, i.e., $\sigma_{x}^{2}=\sigma_{y}^{2}=\sigma_{0}$ in spin space ($\sigma_0$ is the unit matrix in spin space).
On the other hand, they are orthogonal in the latter case, which leads to the off-diagonal component in spin space and results in $\mathrm{Tr}[\cdots]=0$.
Thus, the effective coupling between the spin moment and $g$-vector in the inner-product form is important to lead to the INAHE.

A similar statement can be applied to the AFM state with $T_{z}$.
The lowest-order contribution to $\sigma_{z;xx}^{\mathrm{int, H}}$ arises at $(i,j,k)=(0,1,1)$, where $\Gamma_{z;x,x}^{011}$ is given by 
\begin{equation}
  \Gamma_{z;x,x}^{011}=-16h_{\mathrm{AF}}t_{c}\alpha_{1}(t_{a}^{2}+t_{b}^{2}+4t_{a}^{\prime 2}+4t_{b}^{\prime 2}).
\end{equation}
Similar to the AFM state with $M_{u}$, the conditions of $h_{\mathrm{AF}}\neq 0$, $\alpha_1 \neq 0$, and $t_{c} \neq 0$ are necessary to induce the INAHE.
Meanwhile, the diagonal hopping is not required in contrast to the case with $M_{u}$.
Reflecting such a difference, the contributions of the closed path to $\Gamma_{z;x,x}^{011}$ are distinct, one of which is given by 
\begin{equation}
  \mathrm{Tr}[T_{z}^{(\mathrm{c})}Q_{0}^{(\mathrm{b}1)}Q_{0}^{(\mathrm{c})}Q_{0}^{(\mathrm{b}1)}T_{z}^{(\mathrm{c})}]\neq 0. 
  \label{path-Tz}
\end{equation}
In the case of the AFM state with $T_{z}$, one finds that the coupling between the spin moment and $g$-vector at the same sublattice plays an important role, as schematically shown in Fig.~\ref{path}(c), so as to have its nonzero inner-product.

\section{Summary and Discussion}
\label{summary}

In summary, we have investigated the INAHE in $\mathcal{PT}$-symmetric noncollinear AFMs. 
We have shown that the magnetic toroidal dipole, magnetic quadrupole, and magnetic toroidal octupole contribute to intrinsic nonlinear conductivity.
Especially, we clarified that its Hall part, i.e., INAHE, is accounted for by the emergence of the magnetic toroidal dipole and magnetic quadrupole; the former gives rise to both Ohmic and Hall parts, while the latter induces the pure Hall part.
Based on the microscopic model analysis for the three-dimensional tetragonal lattice models with $M_{u}$ and $T_{z}$, we found two important factors to cause the INAHE: One is the effective coupling between the magnetic order and ASOI and the other is the hopping paths.

Although we have analyzed a specific model in the tetragonal system to demonstrate the INAHE, our result can be straightforwardly applied to other models under different lattice structures once the tight-binding Hamiltonian is provided. 
Furthermore, the effective coupling between the magnetic order and ASOI is a general feature to induce the INAHE in other cases because of the nature of closed paths in Eq.~(\ref{essential}).

Finally, we list candidate materials to have nonzero pure $\sigma_{\mu;\alpha\beta}^{\mathrm{int, H}}$ in accordance with the MPG.
The materials are referred from MAGNDATA \cite{S1600576716012863}, magnetic structure database, in Table \ref{candidates}.
In the listed materials, the origin of $\sigma_{\mu;\alpha\beta}$ is identified as the pure INAHE when $\sigma_{\mu;\alpha\beta}^{\mathrm{int, H}}\neq 0$ but $\sigma_{\mu;\alpha\beta}^{\mathrm{int, O}}=\sigma_{\mu;\alpha\beta}^{\mathrm{D}}=\sigma_{\mu;\alpha\beta}^{\mathrm{BCD}}=0$ for a specific component $\mu,\alpha,\beta$.
Such a situation is satisfied for the MPGs where the magnetic quadrupole $M_{u}$ belongs to the totally symmetric irreducible representation.
It is noted that some MPGs exhibit other contributions from the Drude and BCD terms for different tensor components.
For example, let us suppose the $4m^{\prime}m^{\prime}$, where the BCD term contributes to the conductivity.
In this case, the electric dipole $Q_{z}$ is induced in addition to $M_{u}$~\cite{PhysRevB.104.054412}, but it contributes to only $\sigma_{z;xx}^{\mathrm{H}}$ and $\sigma_{x;zx}^{\mathrm{H}}$, which does not affect $\sigma_{x;yz}^{\mathrm{int, H}}$.
Among the candidate materials, the materials with large spin--orbit coupling and the hybridization between orbitals with different parity are promising, since the ASOI is qualitatively related to them.

\begin{table*}[htbp]
  \caption{\label{candidates}
  Magnetic point group (MPG) and candidate materials to exhibit the pure INAHE of $\sigma_{x;yz}^{\mathrm{int, H}}=-\sigma_{y;zx}^{\mathrm{int, H}}$ except for $\bar{1}^{\prime}$.
  Each row is classified by the existence or absence of Drude and BCD terms for other tensor components based on \cite{PhysRevLett.127.277201}.
  }
  \begin{ruledtabular}
    \begin{tabular}{lcccl}
      $\mathrm{MPG}$ & intrinsic & Drude & BCD & Materials 
      \\
      \hline
      $4/m^{\prime}m^{\prime}m^{\prime}$ & $\bigcirc$ & $\times$ & $\times$ & GdB$_{4}$~\cite{PhysRevB.73.212411}, Fe$_{2}$TeO$_{6}$~\cite{Fe2TeO6}, UPt$_{2}$Si$_{2}$~\cite{UPt2Si2}, Bi$_{2}$CuO$_{4}$~\cite{Bi2CuO4}, UBi$_{2}$~\cite{UBi2}, UGeSe~\cite{UGeSe}
      \\
      $\bar{6}^{\prime}m^{\prime}2$ & & & & RbFeCl$_{3}$~\cite{RbFeCl3}, UNiGa~\cite{UNiGa}, TmAgGe~\cite{TmAgGe}
      \\
      $6/m^{\prime}m^{\prime}m^{\prime}$ & & & & 
      \\
      \hline
      $4m^{\prime}m^{\prime}$ & $\bigcirc$ & $\times$ & $\bigcirc$ & CeCoGe$_{3}$~\cite{PhysRevB.88.134416}, CeIrGe$_{3}$~\cite{PhysRevB.97.184422}
      \\
      $6m^{\prime}m^{\prime}$ & & & & ScMnO$_{3}$~\cite{PhysRevB.62.9498}, HoMnO$_{3}$~\cite{HoMnO3}, LuFeO$_{3}$~\cite{PhysRevLett.114.217602}, Nd$_{15}$Ge$_{9}$C$_{0.39}$~\cite{Nd15Ge9C}, Mn$_{2}$Mo$_{3}$O$_{8}$~\cite{Mn2Mo3O8}, 
      \\
       & & & & YbMnO$_{3}$~\cite{PhysRevB.98.134413}, Yb$_{0.42}$Sc$_{0.58}$FeO$_{3}$~\cite{PhysRevB.103.174102}
      \\
      \hline
      $4/m^{\prime}$ & $\bigcirc$ & $\bigcirc$ & $\times$ & (K,Rb)$_{y}$Fe$_{2-x}$Se$_{2}$~\cite{XFeSe2}, TlFe$_{1.6}$Se$_{2}$~\cite{PhysRevLett.109.077003}, K$_{0.8}$Fe$_{1.8}$Se$_{2}$~\cite{K0.8Fe1.6Se2}, NdB$_{4}$~\cite{JPSJ.86.044705}
      \\
      $\bar{3}^{\prime}$ & & & & MnTiO$_{3}$~\cite{JPSJ.14.1352}, MnGeO$_{3}$~\cite{JPSJ.37.1242}, MgMnO$_{3}$~\cite{PhysRevMaterials.3.124406}, Yb$_{3}$Pt$_{4}$~\cite{PhysRevB.81.064401}
      \\
      $\bar{3}^{\prime}m^{\prime}$ & & & & Cr$_{2}$O$_{3}$~\cite{Cr2O3}, (Co,Mn)$_{4}$Nb$_{2}$O$_{9}$~\cite{PhysRevB.93.075117, PhysRevB.96.094434, PhysRevB.97.085154, PhysRevB.97.020404, PhysRevB.102.174443}, Mn$_{4}$Ta$_{2}$O$_{9}$~\cite{PhysRevB.98.134438, PhysRevB.103.014422}, U$_{2}$N$_{2}$(S,Se)~\cite{U2N2M}, AgRuO$_{3}$~\cite{PhysRevB.103.214413}, 
      \\
       & & & & Na$_{2}$MnTeO$_{6}$~\cite{PhysRevB.105.064416}
      \\
      $\bar{6}^{\prime}$ & & & & Cu$_{0.82}$Mn$_{1.18}$As~\cite{PhysRevMaterials.3.111402}, Tb$_{14}$Ag$_{51}$~\cite{Tb14Ag51}
      \\
      $6/m^{\prime}$ & & & & U$_{14}$Au$_{51}$~\cite{U14Au51}
      \\
      \hline
      $3m^{\prime}$ & $\bigcirc$ & $\bigcirc$ & $\bigcirc$ & U$_{3}$(P,As)$_{4}$~\cite{U3M4}, GaV$_{4}$S$_{8}$~\cite{GaV4S8}, CaBaCo$_{2}$Fe$_{2}$O$_{7}$~\cite{PhysRevB.97.144402}, Tb(DCO$_{2}$)$_{3}$~\cite{TbDCO23}, CrSe~\cite{PhysRev.122.1402} 
    \end{tabular}
  \end{ruledtabular}
\end{table*}

\begin{acknowledgments}
  This research was supported by JSPS KAKENHI Grants Numbers JP21H01037, JP22H04468, JP22H00101, JP22H01183, and by JST PRESTO (JPMJPR20L8).
\end{acknowledgments}

\appendix
\section{Matrix Element of Multipoles}
\label{matrix_element}

We show the matrix elements of the multipoles in the Hamiltonian (\ref{hopping_multipole}), (\ref{asoi_multipole}), and (\ref{mf_multipole}).
The onsite and real bond degrees of freedom are described by $Q^{(\mathrm{c})}$ and $Q^{(\mathrm{b}n)}$ and the imaginary bond degrees of freedom are described by $T^{(\mathrm{b}n)}$.
For the basis $\{\ket{\mathrm{A}\sigma}, \ket{\mathrm{B}\sigma}, \ket{\mathrm{C}\sigma}, \ket{\mathrm{D}\sigma}\}$, the relevant matrix elements of the multipoles are given by 
\begin{equation*}
  \begin{aligned}
    &\, Q_{0}^{(\mathrm{c})}=
    \left(\begin{matrix}
      1 & 0 & 0 & 0
      \\
      0 & 1 & 0 & 0
      \\
      0 & 0 & 1 & 0
      \\
      0 & 0 & 0 & 1
      \\
    \end{matrix}\right)\sigma_{0},
    \ 
    Q_{x}^{(\mathrm{c})}=
    \left(\begin{matrix}
      -1 & 0 & 0 & 0
      \\
      0 & 1 & 0 & 0
      \\
      0 & 0 & 1 & 0
      \\
      0 & 0 & 0 & -1
      \\
    \end{matrix}\right)\sigma_{0},
    \\
    &\, Q_{y}^{(\mathrm{c})}=
    \left(\begin{matrix}
      -1 & 0 & 0 & 0
      \\
      0 & 1 & 0 & 0
      \\
      0 & 0 & -1 & 0
      \\
      0 & 0 & 0 & 1
      \\
    \end{matrix}\right)\sigma_{0},
    \ 
    Q_{0}^{(\mathrm{b}1)}=
    \frac{1}{2}
    \left(\begin{matrix}
      0 & 0 & 1 & 1
      \\
      0 & 0 & 1 & 1
      \\
      1 & 1 & 0 & 0
      \\
      1 & 1 & 0 & 0
      \\
    \end{matrix}\right)\sigma_{0},
    \\
    &\, Q_{v}^{(\mathrm{b}1)}=
    \frac{1}{2}
    \left(\begin{matrix}
      0 & 0 & -1 & 1
      \\
      0 & 0 & 1 & -1
      \\
      -1 & 1 & 0 & 0
      \\
      1 & -1 & 0 & 0
      \\
    \end{matrix}\right)\sigma_{0},
    \\
    &\, Q_{0}^{(\mathrm{b}2)}=
    \left(\begin{matrix}
      0 & 1 & 0 & 0
      \\
      1 & 0 & 0 & 0
      \\
      0 & 0 & 0 & 1
      \\
      0 & 0 & 1 & 0
      \\
    \end{matrix}\right)\sigma_{0},
    \ 
    Q_{xy}^{(\mathrm{b}2)}=
    \left(\begin{matrix}
      0 & 1 & 0 & 0
      \\
      1 & 0 & 0 & 0
      \\
      0 & 0 & 0 & -1
      \\
      0 & 0 & -1 & 0
      \\
    \end{matrix}\right)\sigma_{0},
    \\
  \end{aligned}
\end{equation*}
\begin{equation}
  \begin{aligned}
    &\, T_{x}^{(\mathrm{b}1)}=
    \left(\begin{matrix}
      0 & 0 & -i & 0
      \\
      0 & 0 & 0 & i
      \\
      i & 0 & 0 & 0
      \\
      0 & -i & 0 & 0
      \\
    \end{matrix}\right)\sigma_{0},
    \
    T_{y}^{(\mathrm{b}1)}=
    \left(\begin{matrix}
      0 & 0 & 0 & -i
      \\
      0 & 0 & i & 0
      \\
      0 & -i & 0 & 0
      \\
      i & 0 & 0 & 0
      \\
    \end{matrix}\right)\sigma_{0},
    \\
    &\, T_{x}^{(\mathrm{b}2)}=
    \left(\begin{matrix}
      0 & -i & 0 & 0
      \\
      i & 0 & 0 & 0
      \\
      0 & 0 & 0 & i
      \\
      0 & 0 & -i & 0
      \\
    \end{matrix}\right)\sigma_{0},
    \ 
    T_{y}^{(\mathrm{b}2)}=
    \left(\begin{matrix}
      0 & -i & 0 & 0
      \\
      i & 0 & 0 & 0
      \\
      0 & 0 & 0 & -i
      \\
      0 & 0 & i & 0
      \\
    \end{matrix}\right)\sigma_{0},
  \end{aligned}
\end{equation}
where $\sigma_{0}$ represents the unit matrix in spin space.
The multipole matrices are orthogonal with each other, i.e., $\mathrm{Tr}[X_{p}X_{q}]\propto \delta_{pq}$ for $X=Q$ or $T$.
The subscript $0$, $(x,y)$, and $(v, xy)$ represent the monopole, dipole, and quadrupole components, respectively.

\section{Comparison of Fermi Sea and Fermi Surface Terms}
\label{sea-surface}

\begin{figure}[h]
  \includegraphics[width=\linewidth]{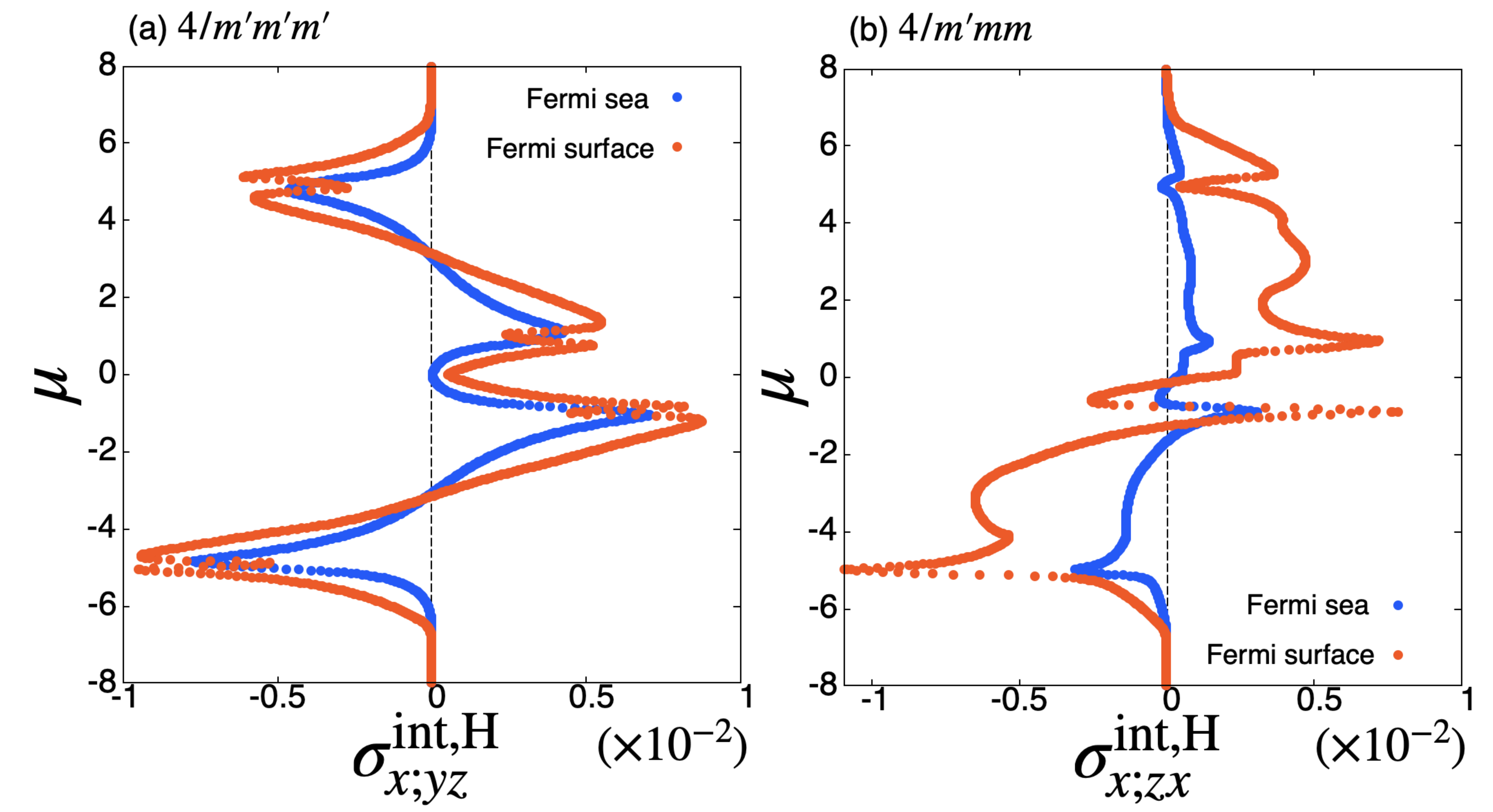}
  \caption{\label{separate}Contributions from the Fermi sea and Fermi surface terms under the AFM states with (a) $M_{u}$ and (b) $T_{z}$ at $T=0.01$.
  We set the same model parameters as Eq.~(\ref{parameters}).}
\end{figure}

We compare the contributions from the Fermi sea and Fermi surface terms in Eq.~(\ref{int}).
In the expression of $\sigma_{\mu;\alpha\beta}^{\mathrm{int, H}}$ in Eq.~(\ref{int-O/H}), the Fermi sea (surface) term corresponds to the first (second) term in the square bracket.
In contrast to the linear anomalous Hall effect, there is no obvious way to transform the Fermi sea term into the Fermi surface term.
On the other hand, the Fermi surface term can be transformed into the Fermi sea term such as
\begin{equation*}
  \begin{aligned}
    &\, \frac{1}{V}\sum_{{\bm k}}\left(-\frac{\partial f_{\nu{\bm k}}}{\partial \varepsilon_{\nu{\bm k}}}\right)
    \frac{v^{\nu}_{\mu}({\bm k})g^{\nu\bar{\nu}}_{\alpha\beta}({\bm k})}{\varepsilon_{\nu{\bm k}}-\varepsilon_{\bar{\nu}{\bm k}}}
    \\
    =&\, -\frac{1}{\hbar V}\sum_{{\bm k}}\partial_{\mu}f_{\nu\bm{k}}\frac{g^{\nu\bar{\nu}}_{\alpha\beta}({\bm k})}{\varepsilon_{\nu{\bm k}}-\varepsilon_{\bar{\nu}{\bm k}}}
    \\
    =&\, \frac{1}{\hbar V}\sum_{{\bm k}}f_{\nu\bm{k}}\partial_{\mu}\left(\frac{g^{\nu\bar{\nu}}_{\alpha\beta}({\bm k})}{\varepsilon_{\nu{\bm k}}-\varepsilon_{\bar{\nu}{\bm k}}}\right).
  \end{aligned}
\end{equation*}
Figure~\ref{separate} shows the $\mu$ dependence of $\sigma^{\mathrm{int, H}}$.
In both cases for the AFMs with $M_u$ [Fig.~\ref{separate}(a)] and $T_z$ [Fig.~\ref{separate}(b)], the behaviors of $\sigma^{\mathrm{int, H}}$, such as the order of the magnitude and their sign dependence, are similar.
Thus, both terms contribute to the INAHE at the same order.
A sudden change of $\sigma^{\mathrm{int, H}}$ is found in the Fermi surface term compared to the Fermi sea term, since the former is more sensitive to the change of the Fermi-surface topology.

\begin{figure*}[t]
  \includegraphics[width=\linewidth]{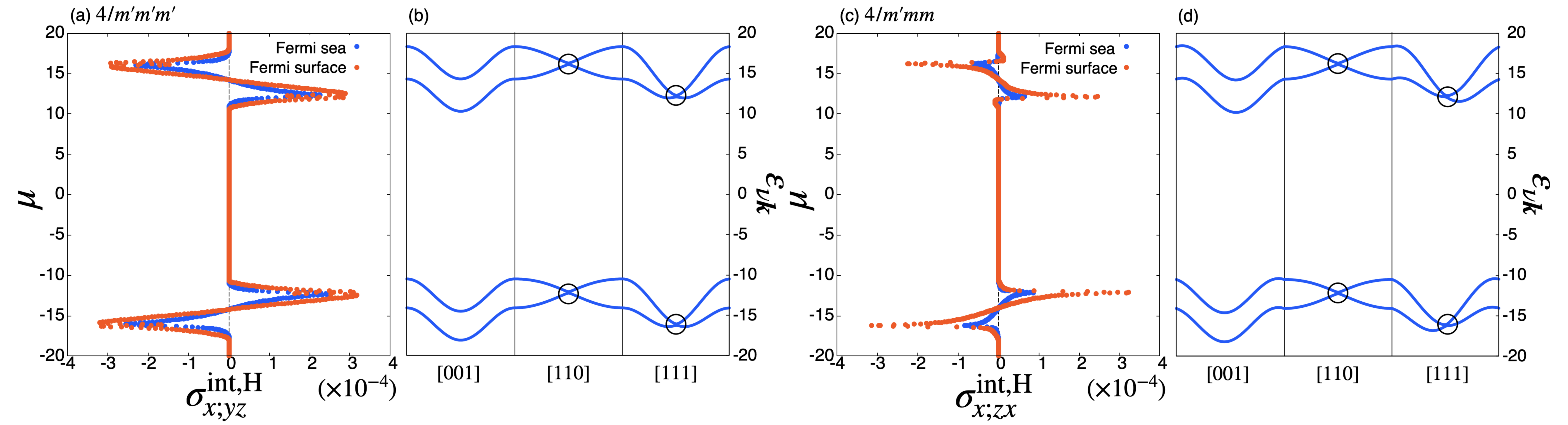}
  \caption{\label{strong}Contributions from the Fermi sea and Fermi surface terms under AFM states with (a) $M_{u}$ under the magnetic point group (MPG) $4/m^{\prime}m^{\prime}m^{\prime}$ and (c) $T_{z}$ under the MPG $4/m^{\prime}m^{\prime}m^{\prime}$ at $h_{\mathrm{AF}}=10$ and $T=0.01$.
  Other parameters are used in Eq.~(\ref{parameters}).
  (b) and (d) show the band structures in each state.
  The circles indicate the narrow band gap region.}
\end{figure*}

We also show the results for a large MF by taking $h_{\mathrm{AF}}=10$ in Fig.~\ref{strong}.
As shown in Figs.~\ref{strong}(b) and (d), the four bands are completely separated into pairs of two bands.
In this case, the contributions from the upper (lower) two bands in $\sigma^{\mathrm{int, H}}$ can be neglected when the Fermi surface is in the lower (upper) two bands in both orderings.
Indeed, comparing Fig.~\ref{strong}(c) with Fig.~\ref{separate}(b), the $\mu$ dependence of $\sigma_{x;zx}^{\mathrm{int, H}}$ at $h_{\mathrm{AF}}=10$ is simpler than that of $h_{\mathrm{AF}}=2$.
The INAHE vanishes at a half-filled region for $-10\lesssim \mu \lesssim 10$ in Figs.~\ref{strong}(a) and \ref{strong}(c).

\bibliography{main_text_v2.bbl}

\end{document}